\documentclass[journal]{IEEEtran}
\ifCLASSINFOpdf
\else
\fi

\usepackage{cite}
\usepackage{amsmath,amssymb,amsfonts}
\usepackage{algorithm}
\usepackage{algpseudocode}
\usepackage[algo2e]{algorithm2e} 

\usepackage{tikz}
\usetikzlibrary{shapes.geometric, arrows}
\usepackage{longtable}

\usepackage{graphicx}
\usepackage{epstopdf}
\usepackage{xcolor}

\newcommand*{\Scale}[2][4]{\scalebox{#1}{$#2$}}%

\newtheorem{thm}{Theorem}
\newtheorem{lem}{Lemma}
\newtheorem{prop}{Proposition}
\newtheorem{cor}{Corollary}
\newtheorem{defn}{Definition}

\newtheorem{rem}{Remark}

\newtheorem{assm}{Assumption}

\usepackage{textcomp}

\usepackage{xargs}                      

\begin{document}

\title{A Risk-Averse Preview-based $Q$-Learning Algorithm: Application to Highway Driving of Autonomous Vehicles}
%


\author{ Majid Mazouchi, Subramanya Nageshrao, and Hamidreza Modares
\thanks{This work was supported by Ford Motor Company-Michigan State University Alliance.}
\thanks{M Mazouchi and H Modares are with  the Department
of Mechanical Engineering, Michigan State University, East Lansing, MI, 48863, USA, (e-mails: mazouchi@msu.edu; modaresh@msu.edu). S. Nageshrao is with Ford Research and
Innovation Center, Ford Motor Company, Palo Alto, CA 94304, USA,
(e-mail: snageshr@ford.com).}
}


%
%

\markboth{}%
{Shell \MakeLowercase{\textit{Mazouchi et al.}}: A Risk-Averse Preview-based $Q$-Learning Algorithm: Application to Highway Driving of Autonomous Vehicles}
%

\maketitle

\begin{abstract}
A risk-averse preview-based $Q$-learning planner is presented for navigation of autonomous vehicles. To this end, the multi-lane road ahead of a vehicle is represented by a finite-state non-stationary Markov decision process (MDP).   A risk assessment unit module is then presented that leverages the preview information provided by sensors along with a stochastic reachability module to assign reward values to the MDP states and update them as scenarios develop.
A sampling-based risk-averse preview-based $Q$-learning algorithm is finally developed that generates samples using the preview information and reward function to learn risk-averse optimal planning strategies without actual interaction with the environment. The risk factor is imposed on the objective function to avoid fluctuation of the $Q$ values, which can jeopardize the vehicle’s safety and/or performance. The overall hybrid automaton model of the system is leveraged to develop a feasibility check unit module that detects unfeasible plans and enables the planner system to react proactively  to the changes of the environment.
Finally, to verify the efficiency of the presented algorithm, its implementation on two highway driving scenarios of an autonomous vehicle in a varying traffic density is considered. 
\end{abstract}



\begin{IEEEkeywords}
Autonomous vehicle, log-expected-exponential Bellman inequality, risk-averse $Q$-learning.
\end{IEEEkeywords}

\IEEEpeerreviewmaketitle

\section{Introduction}
\IEEEPARstart{T}{here}   { are generally two coupled requirements that every autonomous vehicle must satisfy: the vehicle’s performance and its safety.}
Vehicle's performance captures the cost incurred in achieving its goals (i.e., the time of arrival, the energy consumption, the passenger's comfort, etc.). Vehicle's safety requirements impose constraints on its maneuvers to ensure that its trajectories never enter the unsafe set while it traverses toward its goals. The unsafe set is more general than the failure set: the set of states for which a failure, such as a crash, has already occurred. That is, the unsafe set not only includes the failure set but also captures the set of states for which there is no control action to prevent them from eventually entering the failure set.
 { In some real-world applications, the conflict between safety and performance might arise as scenarios develop. 
This is because operating with an optimal performance might increase the vehicle’s risk of getting into an unsafe set in some scenarios (e.g., to have a minimum arrival time, one needs to increase the speed to the maximum allowable speed). This is exacerbated in the presence of uncertainties under which the performance level that can be achieved without safety violation is not known in advance.}
Standard practice typically designs plans that optimize the expected value of vehicle's performance under uncertainties while ignoring its variance.  However, while optimizing the expected value can lead to a good performance on average, not all system responses would be satisfactory. This is because the expected value criterion does not account for the risk of possibility of low-performance trajectories, which can cause an oscillatory behavior around the desired trajectory. This, in turn, increases the risk of getting into an unsafe set. 

Since many sources of uncertainty in autonomous vehicles are aleatory (i.e., random), it is reasonable to design a planner to assure that the cost of its implementation is acceptable for all system trajectories realized from probability distributions of the random sources. Risk-averse strategies have been widely and successfully used to control stochastic systems to avoid performance fluctuation and  reduce the risk of occurrence of catastrophic events, yet low probability \cite{risk2,risk4,risk5,risk6,risk7,risk9,risk11,risk12,risk13,risk14,risk15,risk16,risk17}. Risk measure can be inferred as a mapping from random variables to scalars \cite{doi:10.1137/1.9780898718751,Saldi2020ApproximateME,Buerle2011MarkovDP} and risk-averse optimization interpolates between the expectation-optimization approach and robust worst-case approach by hedging against low-probability events without being overly conservative. Several risk measures are considered in the literature, including Value-at-risk (VaR) measure \cite{jelito2020long,di1999risk,zhu2009worst,chapman2021risk,risk10}  and entropic risk measures \cite{Saldi2020ApproximateME}. The former measure optimizes the expected cost while assuring that its variance is under an acceptable threshold. The latter simultaneously optimizes the average cost and its variance.  Risk-averse planning for autonomous vehicles has also been developed in \cite{ahmadi2020risk} and \cite{safaoui2021risk}. However, they are limited to stationary environments and/or known environments. Besides, they do not leverage the preview information provided by sensors to reduce conservatism. Finally, under epistemic uncertainties (i.e., lack of knowledge), they require direct interaction with the environment to learn about it. However, planning without requiring extensive direct exploration or online actual   interaction is significantly beneficial to autonomous driving, for which online actual interaction might be expensive or dangerous.

Reinforcement learning (RL) \cite{powell2007approximate,sutton2018reinforcement,bertsekas2019reinforcement,kamalapurkar2018reinforcement,lewis2013reinforcement,lewis2012reinforcement,banjac2019data,agarwal2019reinforcement}, as the main tool for solving sequential decision-making problems under epistemic uncertainties, has been widely leveraged to learn an optimal control policy for uncertain systems. To account for both aleatory and epistemic uncertainties, risk-averse RL algorithms have been considered to solve risk-averse stochastic optimal control (RASOC) problems in \cite{fernandez1997risk,chow2017risk,haskell2015convex,haskell2013stochastic,ruszczynski2010risk,shen2013risk,mihatsch2002risk,borkar2002q,prashanth2014policy,tamar2014policy,geibel2005risk,mazumdar2017gradient,dvijotham2014convex}, mainly for Markov Decision Processes (MDPs). Value iteration and policy iteration algorithms are developed in \cite{chow2017risk} and  \cite{ruszczynski2010risk} for risk-sensitive decision-making in an MDP framework. Convex analytic approaches are developed in \cite{haskell2015convex} and \cite{haskell2013stochastic} for risk-aware MDPs. In \cite{shen2013risk}, the authors present iterative risk measures which only depend on the current state rather than on the whole history. In \cite{mihatsch2002risk,borkar2002q}, the authors investigate a risk-sensitive RL algorithm by applying utility functions to the temporal difference error. The most widely used RL algorithm for the RASOC problem is the policy gradient (PG) method \cite{sutton2018reinforcement}. In \cite{prashanth2014policy,tamar2014policy}, the authors employed the PG method to risk-aware MDPs with condition VaR measure appearing in either the objective or constraints. PG has been utilized to learn the solution to VaR setting \cite{geibel2005risk,mazumdar2017gradient}, and entropic utility setting \cite{dvijotham2014convex}. However, PG methods generally result in non-convex optimization problems and are sensitive to fixed gradient step size, an inappropriate choice that can easily make it divergent.  Moreover, existing results on risk-averse RL require online interaction with the environment to learn about the value functions or their gradients.
 {For safety-critical systems, however, it is highly preferred to develop risk-averse RL algorithms capable of learning a policy based only on available data collected from
applying a series of control actions to the system.} 


In this paper, we design a risk-averse high-level planner for the navigation of autonomous vehicles between lanes around static and moving obstacles. A hierarchical control structure is leveraged: A risk-averse high-level planner chooses a sequence of waypoints and their corresponding high-level actions, and a low-level controller drives the autonomous vehicle to assigned waypoints.  From the high-level planner’s perspective, the multi-lane road ahead of a vehicle is represented by a finite-state non-stationary MDP. As scenarios develop, a risk assessment unit (RAU) module leverages preview information of what lies ahead of the ego-vehicle, such as other vehicles and road conditions to map probabilistic reward values to the MDP. The RAU module uses a probabilistic reachability module to calculate the probability of MDP states being occupied by other vehicles or obstacles. The probabilistic reward information of MDP is then obtained based on both probabilities of safety violation as well as a goal-reaching objective. A convex program-based preview-based $Q$-learning planner is then developed that turns the preview probabilistic reward information into a non-iterative risk-averse planning engine. The optimizing-based risk-averse $Q$-learning requires sampling state-action-reward tuples from the MDP. One advantage of the presented approach is that these samples can be generated using the preview reward function without actual interaction with the environment. When new preview information arrives or the infeasibility of the current plan is certified, re-planning is considered. To check for infeasibility, the overall hybrid automaton model of the system is derived to develop the feasibility check unit (FCU) module. The developed scheme is evaluated for a case study of risk-averse autonomous vehicle navigation between lanes around obstacles. The contributions of the paper compared to existing planning approaches are listed as follows:
\begin{enumerate}
\item  Compared to sample-based planning approaches such as rapidly-exploring random tree (RRT) \cite{safaoui2021risk,LaValle2000RapidlyExploringRT,LaValle1998RapidlyexploringRT,Rodrguez2006AnOR} and its variants such as RRT* \cite{Karaman2011SamplingbasedAF}, more realistic objective functions can be optimized by the presented approach, rather than just distance to the goal. 
Moreover, in addition to the expected value, the variance of the cost is also optimized.
Finally, in sharp contrast to RRT, transferring to new scenarios can be performed based on the similarity of the reward functions with past learned scenarios to significantly accelerate the search for an optimal path.

\item 	In sharp contrast to existing (deep) RL algorithms for planning \cite{Ye2021ASO}, the presented approach accounts for the risk or variance of the utility of outcomes which leads to a better trade-off between safety and optimality. Moreover, preview information along with reachability analysis are leveraged to generate imagined samples  {(i.e., preview samples)} without requiring actual interaction with the environment. Utilizing these imagined samples and the FCU module enables the planner to proactively avoid risky exploration of the actual environment and react fast to the changes in the environment. Finally, the presented approaches account for the difficulties of the lower-level controller in traversing between waypoints. 
\end{enumerate}
	
The rest of the paper is organized as follows. Section II provides an overview of the proposed risk-averse learning planner framework and structure. Section III provides the problem statement and formulation. The proposed risk-averse $Q$-learning planner is presented in Section IV. The implementation of the proposed control scheme for two highway driving scenarios of an autonomous vehicle in varying traffic density is given in Section V.  Section VI concludes the paper with a note on our future research.

\textbf{Notation.} Throughout this paper, we use $\mathbb{R}$,  ${\mathbb{R}_{ \ge 0}}$, and $\mathbb{N}$   to denote the sets of real numbers, non-negative real numbers, and non-negative integers, respectively. ${\mathbb{R}^n}$ and ${\mathbb{R}^{n\times m}}$   denote the $n$-dimensional real vector space, and the  ${n\times m}$ real matrix space, respectively. For the given variables $x$ and $y$, $\langle x,y\rangle $ denotes the inner product of  $x$ and $y$. For a random variable  $X$, $\mathbb{E}(X)$ denotes the expectation and  ${\mathop{\rm Var}\nolimits} (X) =\mathbb{E} \left( {{X^2}} \right) - \mathbb{E} {(X)^2}$ refers to the variance of $X$, respectively. $\exp (.)$ is an exponential operator such that  $\exp (x) = {e^x}$. For convenience, ${\mathbb{E}_\zeta }$ refers to the expectation over the random variable  $\zeta$. 




\section{Overview of the Presented Risk-averse Learning Algorithm }

\begin{figure}[!t]
\centering{\includegraphics [width=3.4in] {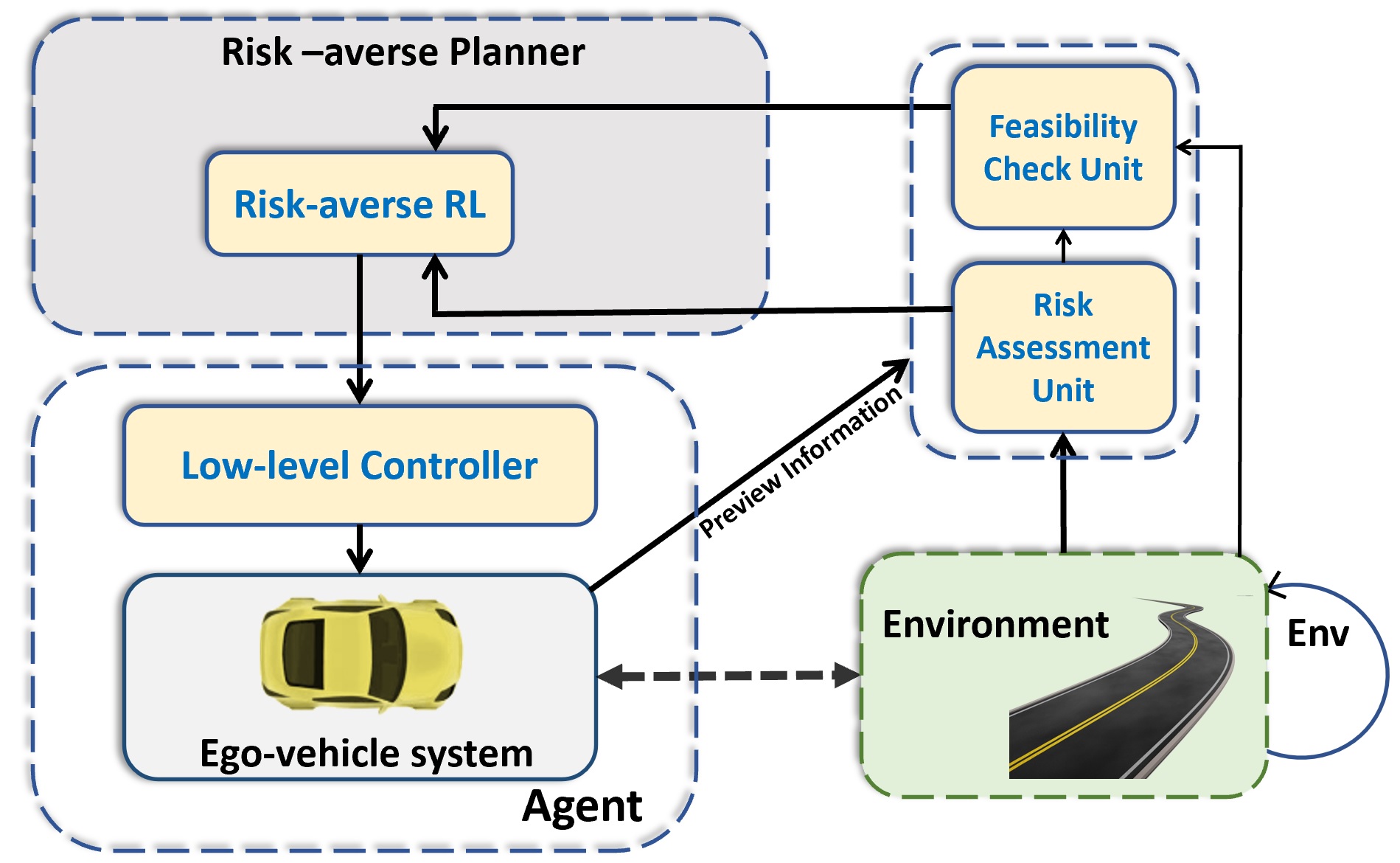}}
\caption{Information flow across modules in the proposed risk-averse planner framework.} \label{Fig1}
\end{figure}

The proposed hierarchical risk-averse control framework is shown in Fig.~\ref{Fig1}. We consider an autonomous vehicle (AV) acting in an uncertain and ever-changing environment. The control system of the AV is hierarchical: A high-level risk-averse learning planner provides a sequence of waypoints to traverse in the discrete-time domain, and a low-level continuous-time controller provides control actuation to drive the AV to the waypoints. The high-level risk-averse planner leverages a preview-based learner in the sense that it learns an optimal plan using the preview information provided by its embedded sensors before a scenario actually becomes apparent. To this end, the RAU module first provides the preview reward function for the current environment. The RAU module is composed of 1) a predictive traffic sub-module that uses embedded sensors along with appropriate sensory fusion and prediction algorithms \cite{Mozaffari_2020} to predict traffic situations, 2) a predictive stochastic reachability sub-module to predict the probability that a traffic participant would occupy a state in the future within a predefined horizon, and 3) a preview reward mapping sub-module to assign reward values to the MDP states. A risk-averse preview-based $Q$-learning planner then turns this preview reward function into a risk-averse waypoint-generation engine. The presented risk-averse $Q$-learning algorithm is a non-iterative optimization-based algorithm for which its samples can be imagined and generated from the AVs’ perspective once the reward function is previewed. 
The risk-averseness accounts for the uncertainty in the reward function (which results from the uncertainty from the behavior of other participants, road conditions, etc.)  as well as the difficulties of the lower-level controller in traversing between waypoints. 
Upon the arrival of new preview information, and before the new scenario becomes apparent, the feasibility check unit (FCU) module uses reachability analysis to verify the feasibility and safety of the planned waypoints along with the corresponding low-level control actuation and triggers re-planning if needed. 
The proposed risk-averse learning planner framework is agnostic to the type of low-level controller and accounts for its difficulty in traversing between waypoints.


\section{Problem Description and Formulation}

\subsection{Agent and Environment Model}

\begin{figure}[!t]
\centering{\includegraphics [width=3.1in] {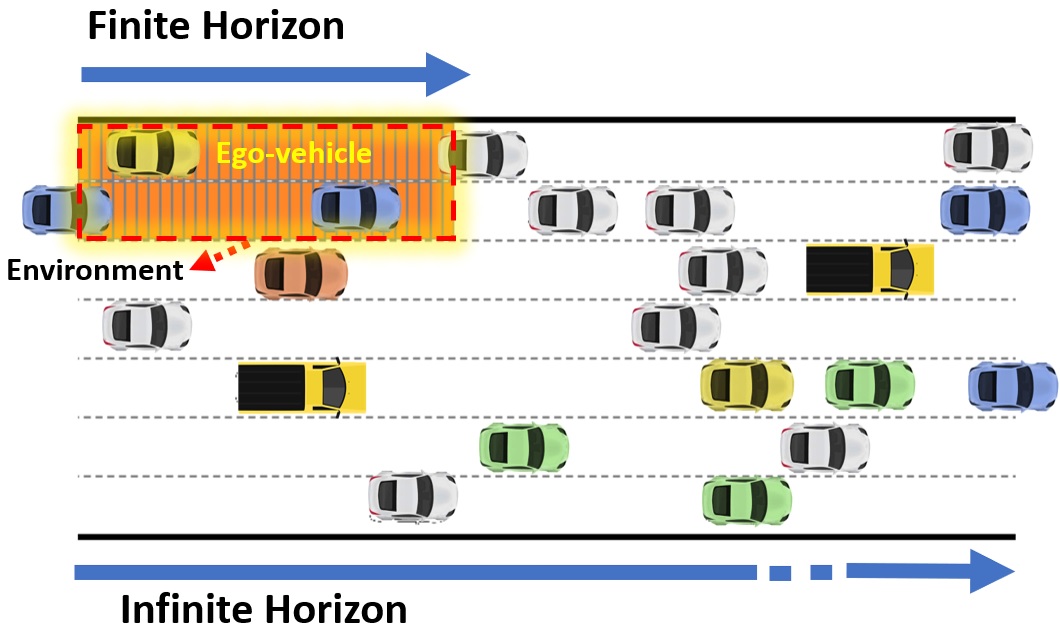}}
\caption{Illustration of an environment as a finite portion of the highway.} \label{Fig3}
\end{figure}

To perform planning in highways, which are actually continuous-state domains, it has become common practice to use discrete planners by uniformly discretizing dimensions of the problem \cite{Schwarting2018AnnualRO,Mueller2017ReinforcementL}. Grid-world Markov Decision Processes (MDPs) are the most common discrete models used for planning. However, planning over a long horizon by finding a policy for the entire domain simultaneously leads to an exponential growth in problem size as the number of states of the MDP increases. To sidestep this problem, the environment (i.e., the ground MDP) is decomposed into a series of grid-world MDPs with a finite number of cells in which the local reward and transition functions are evolved in discrete-time fashion. Therefore, at given predefined discrete-times ${\tau _{env}} > 0$, the AV senses only its neighborhood, i.e., a small part of the highway as shown in Fig.~\ref{Fig3}, and then updates the MDP grid-world (environment) through updating the local reward and transition functions of each cell. It is assumed that the changes in the behavior of other participants and the highway conditions and geometry are insignificant in a small-time horizon ${\tau _{env}}$, and thus the MDP grid-world is assumed to be fixed and stationary for each ${\tau _{env}}$ period of time. 


\begin{figure}[!t]
\centering{\includegraphics [width=2in] {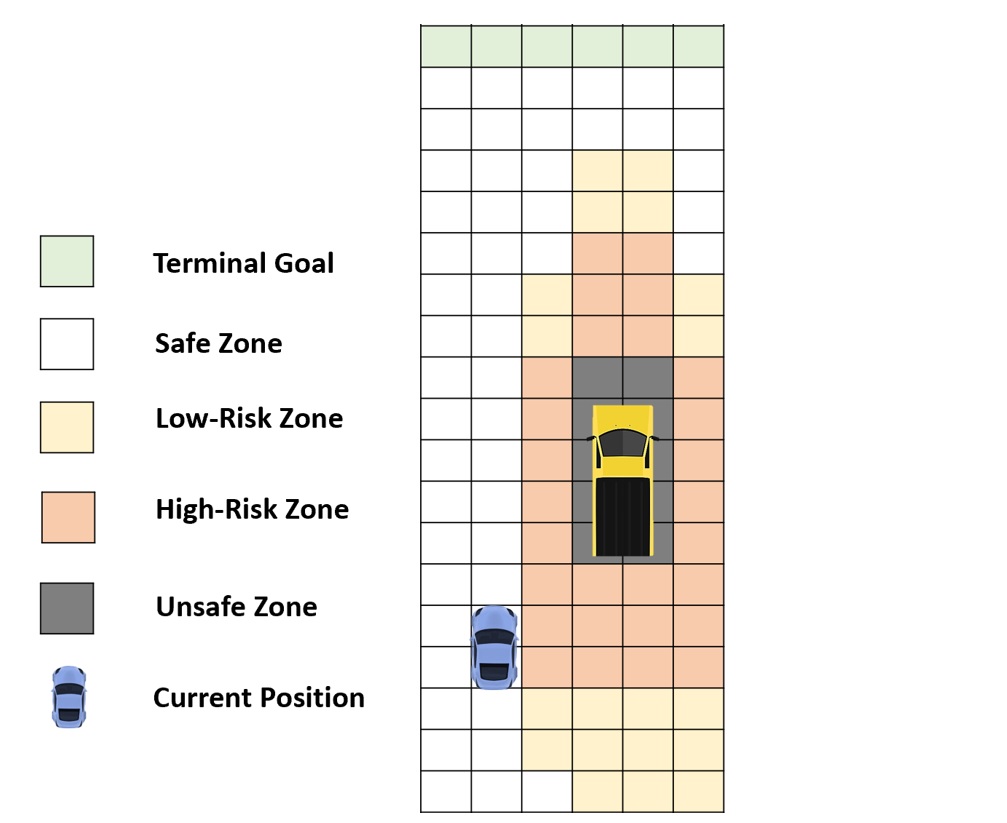}}
\caption{Illustration of the different types of states.} \label{Fig5}
\end{figure}

As scenarios develop, the local rewards corresponding to each cell in the environment are evolved and updated, transitioning the vehicle into a new MDP. The reward function is updated based on both probabilities of goal-reaching objectives and safety violations. Depending on the risk level, which will be determined by the RAU module later, the safety state of each of the $m \times n$  cells in the environment (i.e., the discretized finite portion of the highway around the ego-vehicle) is categorized into the following discrete states ${\cal Z}=\{un,sa,hr,lr,tg,cp\}$ (See Fig.~\ref{Fig5}):
\begin{enumerate}
\item 	{Unsafe $\{un\}$– This kind of state represents the places in the environment that represents failure (i.e., they are occupied with any type of obstacles) or will eventually fail.}
 {\item   Safe $\{sa\}$- This kind of state corresponds to the places in the environment that are not currently in failure states (i.e., Unsafe), and for which there exist control actions that can be taken to avoid them from getting into failure states.}
\item  High-Risk $\{hr\}$- This kind of state represents the places in the environment for which there is a significant chance that the safety mode cannot avoid crashing, i.e., all trajectories starting from these states will end in failure with significant probability.
\item  Low-risk $\{lr\}$- This kind of state represents the places in the environment for which safe mode can avoid crashing, but some possible low-risk actions, such as hard braking or sudden change of direction, are inevitable.
\item  Terminal Goal $\{tg\}$- This kind of state represents the terminal reaching state in the environment.
\item  Current Position $\{cp\}$- This kind of state represents the current position of the AV in the environment.
\end{enumerate}

\begin{figure}[!t]
\centering{\includegraphics [width=3.4in] {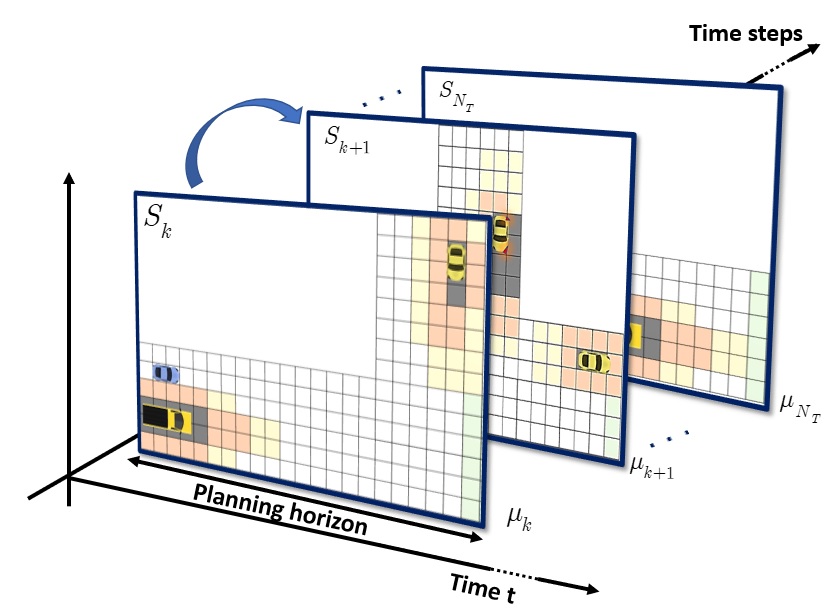}}
\caption{ Illustration of the ground environment MDP ${\cal M}$ as a sequence of consecutive stationary and fixed grid-world MDPs.} \label{Fig3H}
\end{figure}

 {We now formally define the ground MDP ${\cal M}$ which is composed of a series of fixed local MDPs,}
i.e., ${\cal M}=\{{\mu}_1, {\mu}_2,...,{\mu}_{N_T} \}$. The ground MDP captures the transition between local MDPs, as shown in Fig.~\ref{Fig3H}, that are then leveraged by the planner to perform re-planning. The safety states of MPD cells and, consequently, the reward functions of the ground MDP changes from one MDP to another but remain fixed for each local MDP from the planner's perspective. Therefore, the state of each local MDP is limited to the number of cells. 

\begin{defn}\label{defn:10}
\textbf{(Environment ground MDP)}
{The finite-state MDP ${\cal M}$ for the high-level planner consists of a $4$-tuple ${\cal M}:({\cal S}_G,{\cal E},{\cal A},\Pi)$, where:}
{\begin{itemize}
  \item ${\cal S}_G$ is the global state space for the ground MDP with cardinality $|{\cal S}_G| = {[6]^{m \times n}}$, as each cell of each MDP can take six different safety states;  
  \item $\mathcal{E}\equiv\{1,2, \ldots, N_T\}$  is the set of transitory epochs in which local MDPs change, with $N_T \leq+\infty$;
  \item $ {\cal A}$ denotes a finite set of feasible high-level actions;
  \item  $\Pi:=\{\pi_k:~k \in \mathcal{E}\}$ with $\pi_k := \{a_l \in {\cal A}: l=0,1,...\}$, $ k \in \mathcal{E}$, denotes a finite set of feasible policies which maps each global state ${S_k \in {\cal S}_G} $  to a sequence of risk-averse high-level planned actions. 
\end{itemize}}
\end{defn}

Based on Definition \ref{defn:10}, between each consecutive transition epochs, i.e., ${\tau _{env}}$, the environment of AV (i.e., ${\cal M}$) can be seen as a stationary and fixed grid-world MDP. We now define more formally the environment state space for the $k$-th local MDP, $\mu_k$ over a time horizon $(k+1){\tau}_{env}-k{\tau}_{env}$ from the high-level planner's perspective. 



\begin{defn}\label{defn:101}
{\textbf{($k$-th local MDP)} The finite-state local MDP $\mu_k $  consists of a $5$-tuple $\mu_k:({S}_k, {\cal A},  { \bar P_k},{\bar R_k},\gamma)$, where:}
{\begin{itemize}
 \item Finite state space ${S}_k$ with cardinality $|{S}_k|$ where ${S_k}$ consists of $m \times n$ cell states $s$;  
 \item $ {\cal A}$ is a finite set of feasible high-level actions; 
 \item The local transition probabilities
 ${\bar P_k}:= \{ {\bar P_k^l}: l=0,...\}  $ with the conditional probabilities
 ${\bar P}_{s_l,s_{l+1}}^{a} : = {\bar P_k^l}\left( {{s_{l+1} }\mid s_l,a} \right)$,  which denotes the probability of a transition to state  ${s_{l+1} }$ when taking action  $a $ in the state  $s_l$;
 \item The reward function ${\bar R_k }:=\{{\bar R_k^{l}} : l=0,...\}$ where ${{\bar R_k^{l}} := \bar R_k^{l}}({s}_{l},{s}_{l+1},a)$ denotes an immediate reward received by taking action $a \in {\cal A}$ and transitioning from the current state ${s_l} \in {S_k}$ and the next state ${s}_{l+1} \in {S_k}$ based on the reward distribution ${\cal R}_{s_{l},s_{l+1}}^a \in {\cal R}$,  where ${\cal R}$ is the space of all distributions over rewards; 
 \item $\gamma  \in (0,1]$ is the discount factor for future rewards. 
\end{itemize}}
\end{defn}

{ Note that, in the finite-state local MDP  $\mu_k$, ${S}_k$ (resp. ${\cal A}$) is a Borel space (i.e., a Borel subset of a complete and separable metric space) called the state space (resp. action space). For each $s \in {S}_k$ the measurable set ${\cal A}$ denotes the set of feasible high-level actions and has the property that the set of feasible state-action pairs $\{(s, a): s \in {S}_k, a \in {\cal A}\}$ is a measurable subset in ${S}_k \times {\cal A}$.
The canonical sample space is defined as $\Omega:=({S}_k \times {\cal A})^{\infty}$, and all random variables will be defined on the measurable space $(\Omega, \mathcal{B})$ where $\mathcal{B}$ denotes the corresponding product $\sigma$-algebra.}

As an example, one can represent the road ahead as an  $m \times n$ grid-world, where each square (i.e., cell) forward represents the next second of travel, and the squares to the left and right represent lane positions and the edges of the road. For instance, if a vehicle is moving at $100\,kmph$, and each square forward represents one second of the travel, ten  squares represent about $0.35\,km$ in the direction of the travel. Therefore, the environment of vehicle can be modeled as a $2$-dimensional grid-world with $m \times n$ cells over a bounded rectangle (see Fig.~\ref{Fig3}) in which the characteristics of each cell evolve in time.
Moreover, considering $s \in S_k$ as the current state and $s' \in S_k$ as the subsequent state after applying the action $a$, the rewards ${\bar R_k}({s},{s}^\prime,a)$ for this example can be defined as:

\begin{enumerate}
\item Normal distribution with zero mean and very small variance for the high-level action $a$ that cause reaching safe zones, i.e., $sa$. 
\item  Normal distribution with a mean of  $\textbf{M}$ where $\textbf{M} >> 0$ and very small variance for the high-level action $a$ that causes reaching unsafe zones, i.e., $un$.
\item For the high-level action $a$ that causes reaching a next cell state $s^\prime$ with the safety state status $hr$ (i.e., it traverses to a high-risk zone); a random cost can be assigned with truncated normal distribution with the parent normal distribution with a mean of $\textbf{M}_{hr}$ where $\textbf{M}> \textbf{M}_{hr} >> 0$.
\item For the high-level action $a$ that causes reaching a next cell state $s^\prime$ with the safety state status $lr$ (i.e., it traverses to a low-risk zone); a random cost can be assigned with truncated normal distribution with the parent normal distribution with a mean of  $\textbf{M}_{lr}$ where $\textbf{M}_{hr}> \textbf{M}_{lr} > 0$.
\item Normal distribution with a mean of  $\Gamma $  and very small variance 
where $-\Gamma \in \mathbb{R}_{>0}$ is chosen sufficiently large, for the high-level action $a$ that cause reaching a next cell state $s^\prime$ with discrete state terminal goal, i.e., $tg$.
\end{enumerate}

It is assumed here for simplicity that the transition probabilities  ${\bar P_k}\left( {{s^\prime }\mid s,a} \right)$ of the local MDP are known beforehand and not changing over time, i.e., ${\bar P_k}\left( {{s^\prime }\mid s,a} \right)={\bar P}\left( {{s^\prime }\mid s,a} \right)$ for $k=1,...,N_T$.
 {Note that this assumption is reasonable since if the higher-level planner generates a high-level command to move from one waypoint to another one, the low-level controller will follow.}
The probability of transitions can be learned from the success rate of the low-level controller in performing its high-level actions. For example, moving the vehicle from one waypoint to another one when both are located in a straight line is easier for the low-level controller than moving to the left or right with angles more than $42$ degrees. The distributions of the reward functions ${\bar R_k}$, however, change due to the change in the scenario caused by traffic conditions, road geometry, etc. 
 {That is, a rewarding maneuver in the current situation can jeopardize the system’s safety as the scenario develops (e.g., traffic condition changes) and thus will not be rewarding anymore. }
Therefore, as MDPs change in the ground MDP, only their probabilistic reward functions are assumed to change. 

    The planner will learn a policy using stationary and fixed grid-world MDPs, which are called local MDP from now on. This name comes from the fact that the local MDPs of the ground MDP are constructed based on preview information, which can be seen as predicted forms of the actual environment.   
Now, we define the $k$-th local MDP as $\mu_k:({ S}_k, {\cal A},{ \bar P}_k,{ \bar R}_k,\gamma)$, which is used by the high-level planner.
 To take into account future changes in the actual environment that might lead to infeasibility of the planned trajectory, the FCU module is introduced to detect future infeasibility and update the planned trajectory by triggering a re-planning request.

\begin{figure}[!ht]
\centering{\includegraphics [width=3.5in] {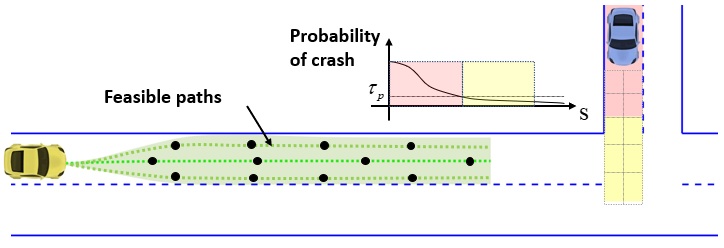}}
\caption{
{Categorizing each cell of an environment based on its probability of crashes. For example, considering the risk threshold of $  0 <  \tau_p <  1$, one can assume that the tiles with probability $(1 - \tau_p)$ are high-risk, and the tiles with probability less than $\tau_p$ are low-risk.}
} \label{cat}
\end{figure}


The objective of the high-level risk-averse planner is to learn the function  ${pl}({{S}}_k):{{\cal {S}}_G}  \mapsto \Pi $ by optimizing a risk-averse objective function with the effective horizon for the local MDP $\mu_k$. That is, the planner learns a sequence of high-level actions ${a_0, a_1,...}$ for the local MDP $\mu_k$, providing the planning policy $\{\pi_k: k \in \mathcal{E}\}$.

\subsection{Risk Assessment Unit and Reward Design Using Preview Information} 

\begin{figure}[!t]
\centering{\includegraphics [width=3.5in] {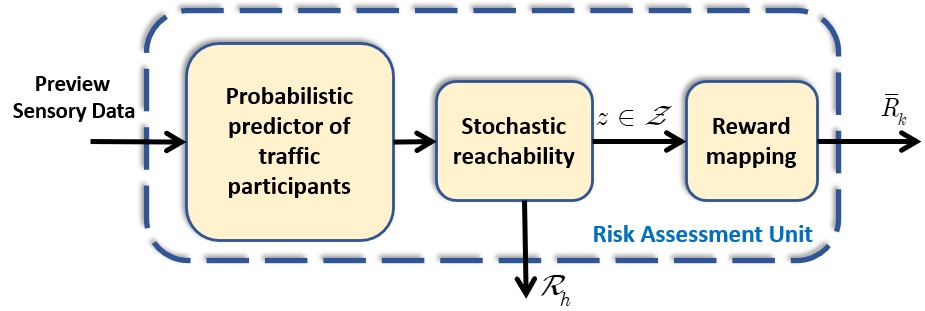}}
\caption{Illustration of the RAU sub-modules: the probabilistic predictor of traffic participants sub-module, the stochastic reachability sub-module, and the reward mapping sub-module.} \label{Fig6}
\end{figure}

An RAU module is developed to design reward functions ${\bar R_k}({s},{s}^\prime,a)$ based on ${\cal R}_{{s},{s}^\prime}^a $ for the $k$-th local MDP, i.e., $\mu_k$, $k \in \mathcal{E}$ as they change over time. 
To learn the reward function that makes a trade-off between safety and performance for each scenario, three modules will be used: a probabilistic predictor of traffic participants, a predictive stochastic reachability module, and a reward mapping module as shown in Fig.~\ref{Fig6}. The probabilistic predictor of traffic participants (PPTP) learns the behavior of traffic participants based on traffic observations provided by sensors such as cameras and LiDAR. The approaches for PPTP include heuristics methods \cite{Brand1997CoupledHM}  and deep learning-based methods for trajectory prediction (e.g., using neural networks \cite{kim2017probabilistic,khosroshahi2016surround}, Bayesian networks \cite{gindele2015learning}), which have shown great performance in predicting the behavior of other traffic participants. 
Using this behavior prediction, which can be as brief as the probability of taking some high-level actions, PPTP predicts the probability of each MDP state being occupied by participants and, thus, its probability of being a failure state. 
However, penalizing possible failure states is not enough for safety guarantee since some states in the MPD are not failure states but are nonetheless high-risk ones in the way that there is a significant chance that the safety mode cannot avoid crashing.
Besides the high-risk zones, there also exist  low-risk zones in which safe mode can avoid crashing with high probability.
Therefore, the stochastic reachability module receives the information of the PPTP module as well as the safety constraints, such as the lane boundaries and the vehicle’s stability limit, etc. It then solves a constraint satisfaction problem over a future time horizon \cite{Bajcsy2019AnER,Liu2019SafetyCW,Althoff2010ReachabilityAA} to 
introduce probabilistic occupancy functions to reason  about  collision  probability and  characterize  keep-out sets for probabilistic safety.
Based on the pre-defined thresholds and the probability of violating safety, each cell then receives a safety state out of the six safety states in ${\cal Z}$ defined before with the corresponding distributions of risk (see Fig.~\ref{cat}).  
These reachable sets are updated in the receding horizon, as new information about the surrounding environment is previewed, e.g., a change in the road curvature or speed of a vehicle is previewed. 
The reward mapping sub-module then utilizes the safety state categorization of cells and the corresponding distribution of risk, as the output of the stochastic reachability module, to assign high reward values for the goal-reaching objective and stochastic penalty values to the actions in each cell state $s$.

Note that the RAU module encodes both safety and performance specifications in the probabilistic reward functions ${\bar R_k}$.
 It is shown in \cite{Massiani2021SafeVF} that a big enough penalty function can assure that the optimal solution for the original task can be learned safely.
 Note also that previewing information is highly advantageous and will allow imagining the reward function ahead of time and thus generate samples for the proposed risk-averse $Q$-learning planner and solve it ahead of time before the scenario becomes apparent.

\section{A Sampling-based Convex Program for Risk-averse preview-based Learning}

\subsection{Risk-averse $Q$-learning for Planning}

Consider the local MDP $\mu_k$ defined based on Definition \ref{defn:101}. 
A policy $\pi_k:{S}_k  \mapsto {\cal A}$ for the high-level planner is a mapping from MDP states $S_k$ to high-level actions ${\cal A}$. Given the local MDP $\mu_k$, the risk-averse planner’s goal is to learn a sequence of high-level actions  to reach a cell state with discrete states type terminal goal, i.e., $tg$, state from an initial state while optimizing a risk-informed cost function. 

Given the finite set of cell states $s \in {S}_k$ and the set of actions $\cal A$, which are generated by the policy $\pi_k$, and the reward function ${\bar R_k }$, all defined in the local MDP $\mu_k$, an infinite-horizon discounted cost is first defined in the form of
 \begin{align}
{{\cal L}^{\pi_k}}({s_0}) = \sum\limits_{l = 0}^\infty  {{\gamma ^l}} {\bar R_k^{l}}. \label{eq:8}
\end{align}
Note that even though the cost is defined over an infinite horizon for simplicity of the analysis and design, the effective horizon of the planning, after which the reward can be ignored, is finite due to the discount factor and depends on the value of the discount factor. Note also that safety and performance specifications are encoded in the reward functions ${\bar R_k^{}}$.

Since the probability transitions ${\bar P}\left( {{s^\prime }\mid s,\pi_k(s)} \right)$ and the reward functions ${\bar R_k^{}}$ provided by the RAU are probabilistic, optimizing the expected value of the cost function (\ref{eq:8}) is not good enough since it can lead to performance fluctuation and even unsafe behaviors. To account for the risk or variance of the performance, which provides high-confidence safety assurance, the entropic-based risk-averse cost function $J$  is used as
\begin{align}
{J_{\pi_k}}\left( s \right) = \frac{1}{\alpha }\log \left( \mathbb{E}{\left[ {\exp \left( {\alpha {{\cal L}^{\pi_k}}} \right)\mid {s_0} = s} \right]} \right), \label{eq:9}
\end{align}
where $\alpha  \in {\mathbb{R}_{ > 0}}$  is the risk-averse factor, and the subscript $l$  is dropped for notational simplicity. 

 {
\begin{lem}\label{Lemma:1n}
Consider the infinite-horizon risk-averse cost function $J_{\pi_k}$ given in (\ref{eq:9}). Then,
\begin{align}
{J_{\pi_k}}\left( s \right) = \mathbb{E} \left( {{{\cal L}^{\pi_k}}\left( s \right)} \right) + \frac{\alpha }{2}{\mathop{\rm \mathbb{V}ar}\nolimits} \left( {{{\cal L}^{\pi_k}}\left( s \right)} \right) + {\cal O} (\alpha^2). \label{eq:10}
\end{align}
\end{lem}}

 {
\textbf{Proof.}  It is well-known that $\exp (f)$ and $\log (1+f)$, respectively, have Maclaurin series $\exp (f)=1+f+{{f}^{2}}/2+ {\cal O}({{f}^{3}})$ and $\log (1+f)=f-{{f}^{2}}/2+ {\cal O}({{f}^{3}})$ which converge for all $f$. Using these facts, substitute the first series into the second one, easily omit terms of higher order than the $3$-th degree denoted by using the big O notation, and some manipulation, then, one has 
\begin{align}
&\log ({\mathbb{E}}(\exp (\alpha {{\mathcal{L}}^{{\pi_k} }}(.))))    =  \alpha {\mathbb{E}}({{\mathcal{L}}^{\pi_k }}(.))+ \nonumber \\ & \left( {{\alpha }^{2}}/2 \right){\mathbb{E}}({{({{\mathcal{L}}^{\pi_k }}(.))}^{2}})
-(1/2){{\left( \alpha {\mathbb{E}}({{\mathcal{L}}^{\pi_k }}(.))+\left( {{\alpha }^{2}}/2 \right) \right)}^{2}} + ... \label{majid1}  
\end{align}  
Expanding the squares, dividing both sides of (\ref{majid1}) by $\alpha $, and using the same reasoning given in \cite{saldi2020approximate}, we have
\begin{align}
&     \frac{1}{\alpha }\log ({\mathbb{E}}(\exp (\alpha {{\mathcal{L}}^{\pi_k }}(.)))) = \nonumber \\ &{\mathbb{E}}({{\mathcal{L}}^{\pi_k }}(.)) 
 +\left( \alpha /2 \right)\left( {\mathbb{E}}({{({{\mathcal{L}}^{\pi}}(.))}^{2}})-{{({\mathbb{E}}({{\mathcal{L}}^{\pi_k }}(.)))}^{2}} \right) + {\cal O} (\alpha^2),
 \label{majid2}
\end{align}   
which completes the proof. $\hfill$ $\square$ 
}


 {It is also shown in \cite{bauerle2018stochastic} that the infinite-horizon risk-averse cost function $J$ given in \eqref{eq:9} can be written as \eqref{eq:10}, which clearly shows that the exponential cost function \eqref{eq:9} takes into account not only the expected value but also the variance of the cost.}


\begin{lem}\label{Lemma:1}
The risk-averse cost function (\ref{eq:9}) is non-decreasing and convex. 
\end{lem}

\textbf{Proof.} See \cite{bauerle2018stochastic}.   \hfill $\square$

We now formally state the entropic risk-averse optimal planning (ERAOP) problem as follows.

\textbf{Problem 1.} Let the reward function (\ref{eq:8}) be previewed by the RAU module based on preview information for a local MDP $\mu_k$. Design a preview-based optimal plan ${\pi_k^*}:{S}_k \mapsto {\cal A}$  that minimizes the risk-averse cost function (\ref{eq:9}) for the resulting MPD $\mu_k$. That is, find  ${\pi_k^*}$ that solves
\begin{align}
{J_{{{\pi_k^*}}}}\left( {{s_0}} \right) = \mathop {\inf }\limits_{{\pi_k} \in {\cal U}} {J_{\pi_k}}\left( {{s_0}} \right) = J_k^*\left( {{s_0}} \right), \label{eq:11}
\end{align}
using only preview information, where ${\cal U}$ is the set of feasible plan policies that assure the boundness of the entropic cost.

 {
\begin{rem} \label{Remark:6}
 It is worth noting that the dynamics of MDP are known only probabilistically, which models the aleatory noise that cannot be reduced as more data are collected. The probabilistic reward functions learned based on the preview information characterize the epistemic uncertainty, which can be determined with more certainty as more rich data are available. In this paper, we assume that the perception system is not under any drift \cite{bitzer2014perceptual} in the performance of perception-based systems, and the uncertainties are caused by unknown behavior of other participants and randomness in traversing between the waypoints. 
 \end{rem}
}


\subsection{Risk-averse preview-based Q-learning }

For a given policy $\pi_k$ and for the cost (\ref{eq:9}), define the value function as
\begin{align}
{V^{\pi_k}}(s): = {J_{\pi_k}}(s). \label{eq:12}
\end{align} 
{
The  $Q$-function associated with the action  $a \in {\cal A}$ at state $s \in S_k$ and following the policy $\pi_k$ afterward is defined as}
\begin{align}
\begin{array}{l}
{Q^{\pi_k}}(s,a): = {g_a^{k}}( s ) + \frac{1}{\alpha } \times \\ \quad \quad \quad  \log \big( {{\mathbb{E}_{s_s^\prime \sim {{\bar P}_{s,s^\prime}^{a}}}}\big[ {\exp ( {\alpha \gamma {V^{\pi_k}}( {s_{a}^\prime })} )} \big]} \big),
\end{array} \label{eq:13}
\end{align}
{where $s_a^{\prime}$ is the realization of the next state starting from the state $s$ and applying the action $a$ and 
\begin{align}
\Scale[1]{ \begin{array}{l}
g_a^k(s): = \frac{1}{\alpha }\log \big( {{\mathbb{E}_{{\cal R}_{s,s'}^a,\,s_a^\prime \sim \bar P_{s,s'}^a}}\left[ {\exp \big( {\alpha \gamma {{\bar R}_k}(s,s',a)} \big)} \right]} \big)\\
\,\,\,\,\,\,\,\,\,\,\,\,\,\,\,\,\, = \frac{1}{\alpha }\log \big( {{\mathbb{E}_{{\cal R}_{s,s'}^a}}\big[ {\sum\limits_{s' \in \,{S_k}} {\bar P_{s,{s^\prime }}^a} \exp \big( {\alpha \gamma {{\bar R}_k}(s,s',a)} \big)} \big]} \big),
\end{array}}
\label{eq:g_a}
\end{align}
$  {\bar P}_{s,s^\prime}^{a}  =  {\bar P}\big( {{s^\prime }\mid s,a} \big) $, and  {${V^{\pi_k}}(s) =  {Q^{\pi_k}}(s,{\pi_k}(s))$}.} 
Defining the optimal $Q$-function ${Q_k^*}(s,a): = {Q^{{{\pi_k^*}}}}(s,a)$
gives the optimal Bellman equation 
\begin{align}
\begin{array}{l}
{Q_k^*}(s,a) = {{g_a^{k}}}\left( s \right) +
\frac{1}{\alpha } 
 \log \big( {{\mathbb{E}_{ s_a^{\prime} \sim {{\bar P}_{s,s^\prime}^{s}} }}\big[ {\exp \big( {\alpha \gamma  {V_k^{*}}\left( {s_a^{\prime} } \right)} \big)} \big]} \big),
\end{array} \label{eq:14}
\end{align}  
where ${a^\prime }$  denotes the next action under the optimal policy ${\pi_k^*}$ and $V_k^{*} = \inf \limits_{{\pi_k} \in {\cal U}} {V^{\pi_k}} = \inf \limits_{{a} \in {\cal A}} {Q_k^*}(s,a)$.  The optimal control policy is then given by
\begin{align}
{\pi_k^*}(s) = \mathop {\arg \min }\limits_{a \in {\cal A}} {Q_k^*}(s,a). \label{eq:15}
\end{align} 
Using the optimal $Q$-risk Bellman operator ${\mathbb{T}}$ as
\small
 \begin{align}
\begin{array}{l}
{{\mathbb{T}}}{Q_k^*}(s,a): = \\
\quad {{{g_a^{k}}}}\left( s \right)  +  \frac{1}{\alpha } 
\log \big( {\mathbb{E}_{s_{a}^\prime \sim {{\bar P}_{s,s^\prime}^{a}}}}\big[ {\exp \big( {\alpha \gamma \mathop {\inf }\limits_{{a^\prime } \in {\cal A}}  {Q_k^*}( {s_{a}^\prime ,a^\prime} )} } \big] \big),
\end{array} \label{eq:16a}
\end{align} 
\normalsize
the optimal Bellman equation becomes
\begin{align}
{Q_k^*}(s,a) = \mathbb{T}{Q_k^*}(s,a). \label{eq:16}
\end{align}

{We use the following definitions in the sequel.}

{\begin{defn}\label{defn:5}
\cite{bertsekas2012weighted} Consider a weight function  $w:{\bf X} \mapsto \mathbb{R}$ with 
$w(x)>0, \quad \forall x \in {\bf X}$
where ${\bf X}$ is a finite space. Let  ${\cal C} ({\bf X})$ the space of real-valued functions ${\cal J}$ on ${\bf X}$ such that ${\cal J}(x) / w(x)$ is bounded as $x$ ranges over ${\bf X}$.
The weighted sup-norm define as
\begin{align}
\|{\cal J}\|_w = \mathop {\sup }\limits_{x \in {\bf X}} \frac{{|{\cal J}(x)|}}{{w(x)}}, \label{eq:19}
\end{align}
on ${\cal C} ({\bf X})$.
\end{defn}}
 
{\begin{defn}\label{defn:4}
\cite{bertsekas2012weighted} An operator ${\mathbb{T}}:{\cal C}\left( {
\bf X} \right) \mapsto {\cal C}\left( {
\bf X} \right)$ is monotone if
\begin{align}
\left\langle {{\mathbb{T}}{{\cal J}_1} - {\mathbb{T}}{{\cal J}_2},{{\cal J}_1} - {{\cal J}_2}} \right\rangle  \ge 0,\;\;\,\forall {{\cal J}_1},{{\cal J}_2} \in {\cal C}\left( {
\bf X} \right). \label{eq:18}
\end{align}
\end{defn}}

 Now to compute the fixed point of \eqref{eq:16}, a convex optimization approach is developed inspired by \cite{de2003linear}. The presented convex optimization approach relies on the Bellman inequality version of \eqref{eq:16}. That is 
 \begin{align}
     {Q_k}(s,a) \le \mathbb{T}{Q_k}(s,a), \label{eq:Bellman_inequality}
 \end{align}
 which implies that ${Q_k} \le {Q_k^*}$ because of the optimal Bellman equation  \eqref{eq:16} and the monotonicity and contractivity properties.
Therefore, to learn the optimal solution to \eqref{eq:16} (i.e., the fixed point of $\mathbb{T}$, one can search for the greatest ${Q_k}$ that satisfies ${Q_k} \le \mathbb{T}{Q_k}$. That is, the following optimization problem solves the Bellman equation \eqref{eq:16}:
\small
\begin{align}
\begin{array}{l}
{\mathop {\sup }\limits_{Q_k \in {\cal C} ({S_k} \times {\cal A})}} \quad \sum\limits_{{S_k} \times {\cal A}} {{{c(s,a) Q_k(s,a)}}}  \\
Q_k(s,a) \le {g_a^{k}}(s)+  \frac{1}{\alpha } \times \\ 
\quad \quad \quad 
 \log \big( {{\mathbb{E}_{s_{a}^\prime \sim {\bar P}_{s,s^\prime}^{a}}}\big[ {\exp \big( {\alpha \gamma  \mathop {\inf }\limits_{{a^\prime } \in {\cal A}} \{Q_k \big( {s_{a}^\prime ,a^\prime )\}} } \big)} \big]} \big)\\
\forall (s,a) \in {S_k} \times {\cal A}.
\end{array}   \label{eq:25vc}
\end{align}  
\normalsize
{where $c(.,.)$ is a non-negative measure (e.g., probability measure \cite{de2003linear}) with finite moments that assigns non-negative mass to all subsets of ${S_k} \times {\cal A}$. It should be noted that $c(.,.)$ is typically used as a probability measure \cite{de2003linear,beuchat2019performance}. For example, a Gaussian distribution can be used when the state-space is unbounded, while a
truncated normal distribution 
or uniform distribution can be used when it is compact. }
The infimum operator in the optimal Bellman operator $\mathbb{T}$ makes \eqref{eq:25vc} hard to solve. It is common practice to remove the infimum operator and instead relax $a'$ in the constraint \eqref{eq:25vc} to for all $ a' \in {\cal A}$.
 This makes the problem convex at the cost of increasing the dimension of constraint.
 Therefore, instead of using the optimal Bellman operator $\mathbb{T}$ and directly solving \eqref{eq:25vc}, we  use an equivalent convex program (will be given later in \eqref{eq:25}) by using the following easier-to-solve Bellman operator, which is developed by dropping the infimum operator in $\mathbb{T}$:
\small
\begin{align}
\begin{array}{l}
\widehat {\mathbb{T}}{Q_k}(s,a): = {g_a^k}\left( s \right) + \frac{1}{\alpha } 
 \log \big(  {\mathbb{E}_{s_{a}^\prime \sim {\bar P}_{s,s^\prime}^{a}}}\big[ {\exp \big( {\alpha \gamma {Q_k}\big( {s_{a}^\prime ,{a}^\prime } } \big)} \big] \big).
\end{array} \label{eq:17}
\end{align}
\normalsize
 
\begin{rem}\label{Remark:New1} 
{Note that the standard policy iteration algorithm cannot be applied to solve the risk-aware optimal control problem \eqref{eq:25vc}. This is because, in contrast to their risk-neutral counterparts, the risk-aware Bellman operator \eqref{eq:16a} is a nonlinear operator, while the standard policy iteration significantly relies on the linearity of Bellman equations. In contrast, the standard value iteration algorithm solves the Bellman equation recursively as
\begin{align}
\begin{array}{l}
{Q_{k+1}}(s,a)= {g_a^k}\left( s \right) + \frac{1}{\alpha } 
 \log \big(  {\mathbb{E}_{s_{a}^\prime \sim {\bar P}_{s,s^\prime}^{a}}}\big[ {\exp ( {\alpha \gamma {Q_k}( {s_{a}^\prime ,{a}^\prime } } )} \big] \big).
\end{array} 
\end{align}
Since the $Q$-function is for the previous iteration on the right-hand side and for the current iteration on the left-hand side, the nonlinearity of the right-hand side does not cause any issue here. That is, this equation is a recursion rather than a Lyapunov equation, and can be easily solved for a nonlinear operator. Therefore, the standard value iteration can also be used to solve the formulated risk-averse problem. Nevertheless, standard value iteration is an iterative approach, and theoretically, it takes an infinite number of iterations for it to converge \cite{bertsekas1996neuro}. In contrast, the presented algorithm here is a one-shot optimization algorithm that needs to be solved only once. Since the infimum operation is removed in \eqref{eq:17}, furthermore, the risk-aware optimal control problem \eqref{eq:25vc} with the relaxed Bellman operator in \eqref{eq:17} can be solved more efficiently than \eqref{eq:16a}, which is used in standard policy iteration or value iteration algorithms.}  
Besides, to learn a risk-averse planning solution for an environment that evolves as a series of MDPs, collecting data for every new MDP is required, which is time-consuming and not practical since a plan must be made before the decision horizon is finished. This is especially true for autonomous driving for which scenarios keep changing, which in turn change the MDP model. The next subsection presents an optimization-based preview-based $Q$-learning that leverages the Bellman operator (\ref{eq:17}) to solve the optimal planning problem in a tractable manner by resolving both non-linearity and non-stationarity issues: The algorithm uses preview information collected by sensors, providing cues about the scenario (e.g., traffic situation, road geometry, etc.), to generate samples of environment interaction without any actual interaction and leverage these samples to generate an optimal plan by solving a static optimization problem before actually, the scenario becomes apparent. 
\end{rem}


 

 Since the presented sampling-based preview-based learning relies on the operator $\widehat {\mathbb{T}}$, it is shown in the following proposition and theorem show that this operator is monotone and contractive.
 
\begin{prop}\label{prop:1}
The Bellman operator $\widehat {\mathbb{T}}$  in (\ref{eq:17}) is monotone.
\end{prop}

\textbf{Proof.} See Appendix A.   \hfill $\square$

We now show that the Bellman operator with a reward function that is not necessarily bounded by a constant is a contraction mapping with respective to a weighted norm defined in Definition \ref{defn:5}. 

\begin{assm}\label{Assm:2}
There exists 
a non-decreasing function $w: \mathbb{R} \mapsto [1,\infty )$  such that $\forall (s,a,{a}^\prime) \in {S_k} \times {\cal A}^2$
\begin{align}
\mathop {\sup }\limits_{a^\prime \in {\cal A}} {\mathbb{E}_{s_a^\prime \sim  {\bar P}_{s,s^\prime}^a}}\{w\left( {s_a^\prime, a^\prime } \right)\} \le \Upsilon w(s,a), \label{eq:21}
\end{align}
for some constant  $\Upsilon  \in \left( {0,{1 \mathord{\left/
 {\vphantom {1 \alpha }} \right.
 \kern-\nulldelimiterspace} \alpha }} \right)$.
\end{assm}

\begin{thm}\label{theorem:1}
Under Assumption \ref{Assm:2}, the Bellman operator in (\ref{eq:17}) is a contraction mapping operator with respect to the weighted norm $w$. That is, 
\small
\begin{align}
\left\|\widehat{\mathbb{T}} {Q_k^2} -  \widehat{\mathbb{T}} {Q_k^1}\right\|_{w} \leq \Upsilon {\gamma}\left\|{Q_k^2} -  {Q_k^1}\right\|_{w}, 
\label{eq:22}
\end{align} 
\normalsize
$\forall {Q_k^1},~{Q_k^2} \in {\cal C} ({S_k} \times {\cal A}) $, where  $\Upsilon \gamma  \le 1$.
\end{thm}

\textbf{Proof.} See Appendix B.   \hfill $\square$

\begin{prop}\label{prop:2}
Let ${Q_k} \in {\cal C} ({S_k} \times {\cal A}) $ be an arbitrary $Q$-function. Then, the following property hold.
\begin{align}
{Q_k^*} = \mathop {\lim }\limits_{l \to \infty } {\widehat {\mathbb{T}}^l}{Q_k}. \label{eq:23}
\end{align} 
\end{prop}

\textbf{Proof.} Proposition \ref{prop:1} and Theorem \ref{theorem:1} show that the Bellman operator satisfies monotonicity and contraction properties. Therefore,
\begin{align}
&\left\|\widehat{\mathbb{T}}^{l}{Q_k}-{Q_k^*}\right\|_{w}=\left\|\widehat{\mathbb{T}} \widehat{\mathbb{T}}^{l-1}{Q_k}-\widehat{\mathbb{T}} {Q_k^*}\right\|_{w} \leq \nonumber \\
&\Upsilon {\gamma}\left\|\widehat{\mathbb{T}}^{l-1}{Q_k}-{Q_k^*}\right\|_{w} \leq \cdots \leq (\Upsilon {\gamma})^l\left\|{Q_k}-{Q_k^*}\right\|_{w}
\end{align}
Now, invoking Banach's theorem \cite{conway2019course} implies that there exists a unique fixed-point solution ${Q_k^{*}}$ of the Bellman operator. This completes the proof.  \hfill $\square$

\begin{prop}\label{prop:3}
A $Q$-function that satisfies the Bellman inequality $Q_k \le  {\widehat {\mathbb{T}}}Q_k$ provides a lower bound to  ${Q_k^{*}}$. 
\end{prop}

\textbf{Proof.} Considering $Q_k \le {\widehat {\mathbb{T}}}Q_k$ and applying the operator ${\widehat {\mathbb{T}}}$  on its both sides repeatedly, and using the results of Proposition \ref{prop:2} yields
\begin{align}
Q_k \le {\widehat {\mathbb{T}}} Q_k \le {{\widehat {\mathbb{T}}}^2}Q_k \ldots . \le {{\widehat {\mathbb{T}}}^\infty }Q_k = \mathop {\lim }\limits_{l \to \infty } {{\widehat {\mathbb{T}}}^l}Q_k = {Q_k^{*}}. \label{eq:24}
\end{align}
This completes the proof.   \hfill $\square$

This proposition is used to develop a preview-based convex problem for learning the risk-averse planning solution using only preview information in a sampling-based optimization framework.  That is, the following preview-based convex optimization is first presented, and its convergence to the fixed point of  $\widehat{\mathbb{T}}$  operator  (\ref{eq:17}) is also shown. Since solving the actual optimization problem can be computationally demanding, a sampling-based finite-horizon convex optimization is then presented, for which the samples are generated using the preview information obtained from the RAU module, to learn a near-optimal solution.


\textbf{Problem 2.} (Preview-based convex program)
\small
\begin{align}
\begin{array}{l}
{\mathop {\sup }\limits_{Q_k \in {\cal C} ({S_k} \times {\cal A})}} 
\quad \sum\limits_{{S_k} \times {\cal A}} {{{c(s,a) } Q_k(s,a)}} 
\\
Q_k(s,a) \le {g_a^{k}}(s)+ \frac{1}{\alpha } 
 \log \big( {{\mathbb{E}_{s_{a}^\prime \sim {\bar P}_{s,s^\prime}^{a}}}\big[ {\exp \big( {\alpha \gamma  Q_k \big( {s_{a}^\prime ,{a}^\prime } } \big)} \big]} \big)\\
\forall (s,a,{a}^\prime) \in {S_k} \times {\cal A}^2
\end{array}   \label{eq:25}
\end{align}  
\normalsize

\begin{cor} \label{Corollary:1}
The solution to the optimization problem (\ref{eq:25}) coincides with the optimal value function ${Q^{*}}$  in (\ref{eq:16}).
\end{cor}

\textbf{Proof.} See Appendix C.    \hfill $\square$

After finding the approximate optimal risk-averse state-action function ${\hat Q_k^{*}}$  from solving Problem 2, one can find the greedy approximate optimal control policy by solving
\small
\begin{align}
\begin{array}{l}
{\pi_k^*}(s) = \mathop {\arg \min }\limits_{a \in {\cal A}} \big\{{\hat Q_k^{*}} (s,a) \big\}.
\end{array} \label{eq:26}
\end{align} 
\normalsize
Note that the size of the decision variables is finite and is equal to the number of sate-action pairs. The number of constraints is also finite since the number of states and actions of MDP are finite. However, to avoid requiring any online interaction to form the constraints, and to avoid computational intractability, the preview reward function as the output of the RAU module is leveraged to generate samples. Note that the transition probabilities for the high-level planner are known and depend on the difficulty of the low-level controller in executing and traversing between waypoints. Therefore, optimization is formed using only preview information and solved for the MDP ahead before the scenario actually becomes apparent.  Algorithm 1 provides the details of the risk-averse preview-based $Q$-learning algorithm.


\begin{algorithm}[!htp] 
\caption{One-shot preview-based risk-averse $Q$-learning algorithm. }
\textbf{Result:} The approximate optimal risk-averse state-action values ${Q_{k}}$.\\
\textbf{Data collection:} \\
 $\null$    $\quad$  \textbf{\textup{for}} $i = 1,..., |S_k|$ \textbf{do} \\
 $\null$    $\quad$ \,\,
 \textbf{\textup{for}} $j = 1,..., |{\cal A}|$ \textbf{do} \\
$\null$    $\quad$ \,\,\,\,\, \textbf{\textup{for}} $l = 1,..., N_l$ \textbf{do} \\
$\null$    $\quad$ \,\,\,\,\,\,\,\,\,
Obtain $(s_i, a_j, {\mathbb S}_{i,j}^\prime, \bar {\mathbb R}_{i,j}^\prime, g_{a_j}^{k}(s_i))$ \\
$\null$    $\quad$ \,\,\,\,\,\,\,\,\, \quad \quad \quad ${\mathbb S}_{i,j}^\prime := \{s_{i,a_j}^{\prime, l}, l=1,...,N_l \}$ \\
$\null$    $\quad$ \,\,\,\,\,\,\,\,\, \quad \quad \quad 
$\bar {\mathbb R}_{i,j}^\prime := \{\bar R_k({s_i},{s}_{a_j}^{ \prime, l},a_j), l=1,...,N_l \}$ \\
$\null$    $\quad$ \,\,\,\,\,\,\,\, \,
Create 
\begin{align}
    {\bar {\cal D}^k} :=  \big\{(s_i, a_j, {\mathbb S}_{i,j}^\prime, \bar {\mathbb R}_{i,j}^\prime, g_{a_j}^{k}(s_i))\big\}_{i=1,j=1}^{|{ S_k}|,|{\cal A}|}
\end{align}
$\null$    $\quad$ \,\,
Select $(s_i, a_j, {\mathbb S}_{i,j}^\prime, \bar {\mathbb R}_{i,j}^\prime, g_{a_j}^{k}(s_i)) \in {\bar {\cal D}^k}$
\\
$\null$    $\quad$ \,\,
Select $a_z^{\prime} \in {\cal A}$
\\
$\null$    $\quad$ \,\,
Create 
\begin{align}
    {{\cal D}^k} :=  \big\{(s_i, a_j, {\mathbb S}_{i,j}^\prime, \bar {\mathbb R}_{i,j}^\prime, g_{a_j}^{k}(s_i),a_z^\prime)\big\}_{i=1,j=1,z=1}^{|{ S_k}|,|{\cal A}|,|{\cal A}|}
\end{align}
\textbf{Function evaluation:} \\
 $\null$    $\quad$ \,\, Solve 
\small
\begin{align}
\begin{array}{l}
\mathop {\sup }\limits_{Q \in {\cal C} ({S_k} \times {\cal A})} \sum\limits_{{s_k} \times {\cal A}} {\rm{c}} ({\rm{s}},{\rm{a}}){{Q}}({{s}},{{a}})\\
     Q( {s_i},a_j ) \le {g_{a_j}^k}( {{s_i}} ) + 
\frac{1}{\alpha }\log (
{\mathbb{E}_{{s_{i,a_j}^{\prime}} \sim {\bar P}_{{s_i},{s_i^{\prime}}}^{a_j}}} 
[\exp (\alpha \gamma   
   Q( {s_{i,a_j}^{ \prime }},a_z^{\prime} ]) \\
\forall (s_i, a_j, {\mathbb S}_{i,j}^\prime, \bar {\mathbb R}_{i,j}^\prime, g_{a_j}^{k}(s_i),a_z^{\prime}) \in {{\cal D}^k}
\end{array}
\label{eq:27}
\end{align}
\normalsize
\end{algorithm} 

\begin{rem}\label{Remark:1}
{
Even though not considered in this paper, it has the potential to leverage transfer learning  based on similarity measured developed for MDPs \cite{lazaric2011transfer,ammar2014automated,visus2021taxonomy} to quickly warm-start Algorithm 1 for an unseen MDP or even using  feature-based exploratory mechanisms such as \cite{vallon2020data}. Transferring of samples can also be performed for the portion of the state space for which the occupancy remains unchanged from one local MDP to another.} \end{rem}

\section{Feasibility Check Unit}


Upon planning high-level actions for the next $n{\tau}_{env}$ seconds at the transition epoch $k$ by the high-level $Q$-learning planner based on the local MDP $\mu_k$, a desired continuous-time reference trajectory for low-level control to track is computed for the next $n{\tau}_{env}$ seconds.
To continuously monitor the feasibility of the generated desired reference trajectory for the next $n{\tau}_{env}$ seconds of the travel, which can be jeopardized due to the change in environment,
 a feasibility check unit (FCU) module is utilized. If the FCU module infers that feasibility violation is imminent at any time in the next $n{\tau}_{env}$ seconds with a probability above a threshold, it triggers a re-planning request. Re-planning can be reactive to just get around infeasible areas and proactive by triggering the preview-based risk-averse $Q$-learning with new information to re-learn a new temporally risk-averse plan.

Since the FCU module needs to consider both  continuous-time transitions (low-level controller outcome) of the AV and discrete-time stochastic transitions of the environment MDP $\cal M$, a hybrid automaton is used to formalize the characteristics of the FCU module.


\begin{figure}[!t]
\centering{\includegraphics [width=3.4in] {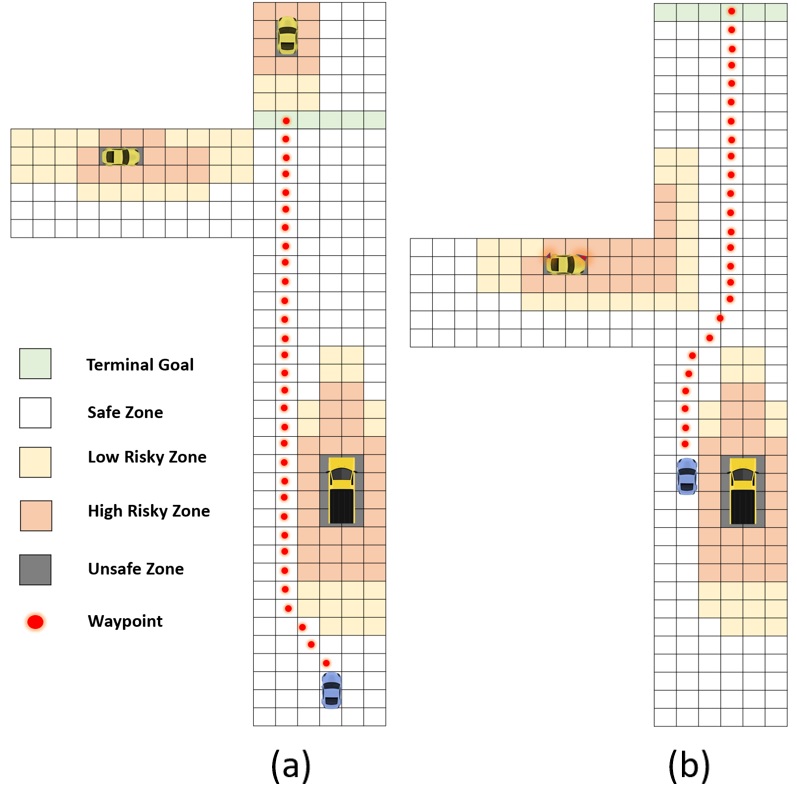}}
\caption{Re-planing due to an infeasibility issue based on preview information. } \label{Fig7}
\end{figure}

Before formally defining the hybrid automata, let us first recall and define some update functions acting on different global states to produce different layers of discrete transitions based on the environment ground MDP ${\cal M}$. Recall that the planner function $pl(.)$ maps each global state ${S_k \in {\cal S}_G} $ in the ground MDP $\cal M$ to a sequence of risk-averse high-level planned actions $a \in {\cal A}$. 
We now define the following functions ${\mathbf{AE}}(.)$ and ${\mathbf{EN}}(.)$ that predict the changes in MDPs based on a change in the actions of the AV itself and a change in the surrounding environment, respectively. These functions are only defined to define the hybrid automata and the presented approach does not need to learn them or know them.
\begin{itemize}
  \item The update function  ${\mathbf{AE}(S_k,pl(S_k))}:{\cal S}_G \times {\cal A} \mapsto {\cal S}_G$ that captures the evolution of local MPDs within the ground MDP $\cal M$  after the AV applies a high-level planned action $a \in {\cal A}$ based on the planner function $pl(.)$. 
  \item The update function $\mathbf{En}(S_k): {\cal S}_G \mapsto {\cal S}_G$ that captures the evolution of local MPDs within the ground MDP $\cal M$ for a future finite-time horizon and based on environmental changes. 
\end{itemize}

Note that, based on the $pl(.)$ function and environmental changes, the $\mathbf{En}(.)$  and ${\mathbf{AE}(.)}$ functions predict and calculate the environment behavior and conditions, such as new locations of the other vehicles  for a future finite-time horizon in the vision horizon of the AV and also predict and calculate the corresponding safety state of zones, new terminal goal, and position of the AV. For notational simplicity in the sequel, the subscript $k$ is dropped when it is clear from the context.


The hybrid automaton is now defined as a $4$-tuple ${\cal H} = \langle {\cal Q},\Theta , \Pi, {\cal T} \rangle $  where  ${\cal Q} = {{\cal S}_G \times {\cal S}_o} \times \Delta $ is the hybrid state space of the automaton, $\Pi$ denotes the set of all possible planners,
$\Theta$ is the initial state, and ${\cal T}$  is a set of continuous-time trajectories. Moreover, ${\cal S}_o$ is the set of vehicle states from the low-level controller perspective which can change continuously, and $\Delta  := \{  + ,\neg  \} $ where  $ + $ indicating that a sequence of planned high-level actions is not feasible anymore as scenarios develop and $\neg $ indicating that a sequence of high-level actions is feasible. Let  $q \in {\cal Q}$ denote a hybrid automaton system state. We denote the components of  $q$ as  $q \cdot {\cal  S}_o$,  $q \cdot S$,  $q.\Delta $.  Moreover, $\Theta  \subseteq {\cal Q}$  is the set of initial states. A trajectory   $\xi  \in {\cal T}$ is a map from  ${\cal Q}$ to  ${\cal Q}$ for a finite time interval $t \in {\mathop{\rm dom}\nolimits} (\xi )$  based on the corresponding high-level planned actions. For the entire time period of the trajectory $\xi_{l,k}(t)  \in {\cal T}$, i.e., $\tau_{env}$, the global environment state $S \in {\cal  S}_G$  remains unchanged. That is, $\forall t \in \tau_{env}$, $\xi_{l,k} (t) \cdot S = \xi_{l,k} (0) \cdot S$. However, $\xi_{l,k} (t).{S_o}$ where $S_o \in {\cal S}_o$ changes continuously with time and is based on the interaction of the AV and low-level controller dynamics. For the sake of simplicity, in this paper, $\xi_{l,k} (t).{S_o}$  is assumed to be a solution of a known differential equation (i.e., the autonomous overall low-level closed-loop system) as follows
\begin{align}
\frac{d}{{dt}}\left( {\xi_{l,k} (t) \cdot {S_o}} \right) = {f_{{q_0} \cdot a}}\left( {\xi_{l,k} (t) \cdot {S_o}} \right),\,\,{q_0} \in {\cal Q}. \label{eq:7}
\end{align}

\begin{assm}\label{Assm:1}
Let the high-level planner action ${a} \in {\cal A}$ commands the vehicle to move from the cell state $s \in S_k$ to $s^\prime \in S_k$. Then, the AV has a low-level controller that can track a  trajectory  $\xi_{l,k}  \in {\cal T}$ in finite time  $\tau_{env}$ such that $\xi_{l,k} (0).{S_o}=s$ and $\xi_{l,k} (\tau_{env}).{S_o}=s^\prime$.

\end{assm}

Three types of discrete transitions are considered: environment transition, risk-averse planner transition, and re-planning transition, denoted by the finite-state set  ${{\cal D}_{Tr}} = \{ env,rpl,pl\}$. Let ${\cal D} \subseteq {\cal Q} \times {{\cal D}_{Tr}} \times {\cal Q}$  denote the set of discrete transitions. Let  $q \in {\cal Q}$ and ${q^\prime } \in {\cal Q}$  denote a state and the post-state of the transition updates of the hybrid automaton system  ${\cal H}$. Let  ${\rm{\bf Trans}}\left( {q,\ell ,{q^\prime }} \right)$ be a discrete transition map between $q$ and $q^\prime$ where  $\ell  \in {{\cal D}_{Tr}}$. Then, these discrete transitions are defined as follows:

\begin{enumerate}
\item   (Environment transition, $\ell  = env$). If the time elapsed since the last environment transition is equal to ${\tau _{env}}$,  then $\ell  = env$  and the environment transition is enabled, i.e., ${\rm{\bf Trans}}\left( {q,\ell  = env,{q^\prime }} \right) \in {\cal D}$, and as a result, the discrete state of the environment set is updated according to the environment update function $\mathbf{En}(.)$, i.e.,  $q \cdot {S_k} \mapsto {q^\prime }.{S_k}$,  and  ${q^\prime } \cdot {S_o} = q \cdot {S_o}$. Moreover, the agent (i.e., the low-level closed-loop system) and the current environment states get updated according to the joint AV/environment update function  ${\mathbf{AE}(S,pl(S_k))}$, i.e., $\left( {{q^\prime } \cdot {S_o},{q^\prime } \cdot {S_k}} \right) = \mathbf{AE}( q \cdot {S_o},q \cdot {S_k})$. 

\item   (Risk-averse planning transition,  $\ell  = pl$). If, at the state  $q$, the time elapsed since the last plan transition is equal to  ${\tau _{pl}}= n{{\tau _{env}}}$ and $q.\Delta  =  \neg  $, then  $\ell  = pl$ and the risk-averse planner transition is enabled.  As a result,  the risk-averse planner $Q$-learning  $pl(.)$ gets updated by the risk-averse planner based on the previewed probabilistic reward information provided by the RAU module. 
\item (Re-planning transition,  $\ell  = rpl$). If, at the state  $q$, $q.\Delta  =  + $, which indicates that the current sequence of waypoints along with their corresponding trajectories is not feasible anymore and further updates for the risk-averse planner are needed, then $\ell  = rpl$  re-planning transition is enabled. The agent (i.e., the low-level closed-loop system) and environment states get updated and then the safety mode controller overrides the current desired planned trajectory as ${q^\prime } \cdot {\xi_{safe}} = \mathbf{SM} \left( {q \cdot {S_k},q \cdot {S_o}} \right)$ for a predefined period of time $\tau_{safe} \ge \tau_{env}$ where $\mathbf{SM}(.)$ is the safety mode controller and then flag $\ell  = pl$. 
\end{enumerate}



\begin{defn}\label{defn:1}
A closed execution fragment of ${\cal H}$ is an alternating finite sequence of finite-domain trajectories and transition labels,  $\eta  (t) = {\xi _0}{\ell _1}{\xi _1}{\ell _2} \ldots {\xi _n}$ such that $\eta  .istate = {\xi _0}.istate \in \Theta $  where $istate$ stands for initial state at each time frame, each  ${\xi _l} \in {\cal T}$, ${\ell _l} \in {\cal D}_{Tr} $  and $\left( {{\xi _l}.} \right.fstate,{\ell _{l + 1}},{\xi _{l + 1}}.istate) \in {\cal D}$ where $fstate$ stands for final state. The last state of a closed trajectory $\eta (t)  = {\xi _0}{\ell _1} \ldots {\ell _{n-1}}{\xi _n}$, is the final state of the last trajectory ${\xi _n}$ in $\eta  $, that is,  $\eta  .fstate = {\xi _n}.fstate$. The duration of a closed execution fragment  $\eta  (t)$, denoted by  $\eta  .dur \le n{\tau_{env}} $, is the sum of the duration of all its trajectories.
\end{defn}

\begin{defn}\label{defn:2}
If there exists a closed execution fragment $\eta $  within a finite time $\eta .dur = {\tau _\eta  }$ where $\eta  .istate \in {{\cal Q}_0} \subseteq {\cal Q}$  and $\eta  .fstate = q \in {\cal Q}$, then the state $q$  is called reachable from the set of states ${\cal Q}_0$.  Moreover, ${{\mathop{\rm \mathbf{Reach}}\nolimits} _{\cal H}}\left( {{Q_0},{\tau _\eta  }} \right)$ denotes the set of all reachable states from  ${\cal Q}_0$. 
\end{defn}

Let ${{\cal R}_h} \subseteq {S_o} \times {S}$  denote a set of high-risk and unsafe hybrid states where ${{\cal R}_h}$ is updated periodically by the RAU module.   

\begin{defn}\label{defn:3}
It is defined that a closed execution fragment $\eta  (t) = {\xi _0}{\ell _1}{\xi _1}{\ell _2} \ldots {\xi _n}$  is feasible during ${\tau _\eta  }$ with respect to ${{\cal R}_h}$, if  $\left( {{{{\mathop{\rm \mathbf{Reach}}\nolimits} }_{\cal H}}({q_0}) \cdot {S_o},{{{\mathop{\rm \mathbf{Reach}}\nolimits} }_{\cal H}}({q_0}) \cdot {S}} \right) \cap {{\cal R}_h} = \emptyset$.
\end{defn}

The FCU module at each time ${\tau_{env}} >> {\tau _{FCU}} > 0$ monitors the environment with respect to the last updated set of high-risk and unsafe hybrid states ${{\cal R}_h}$, which is provided with the RAU module, and current closed execution fragment  $\eta  (t) = {\xi _0}{\ell _1}{\xi _1}{\ell _2} \ldots {\xi _n}$, to detect any infeasibility issues for duration time  $\eta .dur$. If the FCU module detects any infeasibility issue, it is updated $q.\Delta $  as  $ + $. See Fig.~\ref{Fig7}.

 {
\begin{rem}\label{Remark:4}
It is worth noting that after the re-planning request, the developed data-based $Q$-learning algorithm provides a new optimal risk-averse solution based on the updated local MDP provided by the RAU module based on the updated preview sensory data from the current actual traffic scenario evolution. This fact will guarantee the planned trajectory stays optimal risk-averse during actual operation.
It is also worth mentioning that the design parameter $n$ is related to the computational burden of the developed algorithm. 
For example, for $n=1$ (i.e., very short future time horizon),
we need to re-plan very fast with the frequency of $1/{\tau _{env}}$, which increases our computational cost, and if one chooses $n>>1$ to decrease the computational cost, then we might lose some optimality. However, for most traffic scenarios, the flow of traffic is smooth enough (e.g., drivers do not do sudden lane changing or break), and to decrease the computation burden, $n$ can be reasonably increased. 
\end{rem}}

 {
\begin{rem}\label{Remark:5}
It is worth noting that \cite{vallon2020data} provides a data-driven hierarchical learning architecture for predictive control in unknown environments. 
In \cite{vallon2020data}, the authors first design the target sets from past data and then incorporate them into a model predictive control scheme with a shifting horizon that ensures the safety of the closed-loop system when performing the new task. Even though adaptation to the situations for model predictive control (MPC) is considered in \cite{vallon2020data}, which is promising, in contrast to this work, the constraints are adapted in contrast to the reward function. 
Risk-averse MPC is also reported in \cite{ning2021online,sopasakis2019risk,sajadi2017risk,chow2014framework,schuurmans2020learning}. An online learning-based stochastic MPC framework is proposed in \cite{ning2021online}, in which CVaR constraints on system states are required to hold for a family of distributions called uncertainty sets. Based on generic cost functions, the author derives Lyapunov-type risk-averse stability conditions for MPC problems with constrained nonlinear Markovian switching systems in \cite{sopasakis2019risk}. A stochastic nonlinear model predictive control approach was developed in \cite{sajadi2017risk} in order to systematically find an optimal decision for dynamical systems involving uncertainty. An MPC framework for linear systems with multiplicative uncertainty is provided in \cite{chow2014framework}. While MPC results in a non-convex optimization problem for nonlinear systems and/or non-quadratic costs (which are the cases of the risk-aware cost), our risk-aware approach results in a convex optimization approach. Moreover,  MPC with a fixed cost function cannot guarantee a viable solution in ever-changing environments for which the reward function must be adjusted to the situation to trade off safety and optimality.
\end{rem}}

\section{Simulation}

 {As case studies to show the validity of the proposed risk-averse planner scheme, two traffic scenario examples are considered. 
\\
\textbf{Example 1}. 
In the first example, which is the  highway driving of an autonomous vehicle in a varying traffic density, the main objective of the autonomous vehicle (so-called ego-vehicle) is to avoid unsafe and risky zones as they appear while optimizing a goal-reaching objective. In this scenario, for simplicity, we assume that the speed of the vehicles and all other vehicles are constant.}


The model dynamics of ego-vehicle is chosen as a single-track bicycle model vehicle given by
\begin{align}
\begin{array}{l}
\left[ \begin{array}{l}
\,\dot Y\\
\,\dot \Psi \\
{{\dot \alpha}_T}\\
\,\ddot \Psi 
\end{array} \right] = \left[ {\begin{array}{*{20}{c}}
0\\
0\\
{\frac{K}{{{m_T}{v_{T}}}}}\\
{\frac{{1.6154K}}{{{I_T}}}}
\end{array}} \right]\delta  + \\
\left[ {\begin{array}{*{20}{c}}
0&{{v_{T}}}&{{v_{T}}}&0\\
0&0&0&1\\
0&0&{ - \frac{{2K}}{{{m_T}{v_{T}}}}}&{ - \frac{{{m_T}{v_{T,0}} + \frac{{{\rm{ - 0}}{\rm{.2692}}K}}{{{v_{T}}}}}}{{{m_T}{v_{T}}}}}\\
0&0&{\frac{{{\rm{0}}{\rm{.2692}}K}}{{{I_T}}}}&{ - \frac{{{a^2}K + {b^2}K}}{{{I_T}{v_{T}}}}}
\end{array}} \right]\left[ {\begin{array}{*{20}{c}}
Y\\
\Psi \\
{{\alpha_T}}\\
{\dot \Psi }
\end{array}} \right],
\end{array} \label{eq:46t}
\end{align}
where  $Y(t)$  and $\delta (t)$  denote the vertical position and steering angle of the controlled vehicle along its lane, respectively. Moreover, $\Psi $  and $\dot \Psi $ are the yaw and yaw rate, respectively,  ${\alpha _T}$ is the vehicle sideslip angle, $\dot{X}=v_{\mathrm{T}}+X(0)$ and $\dot{v}_{\mathrm{T}}=0$, where $X$ and $v_{T} $ are the longitudinal position of the controlled vehicle and vehicle velocity, respectively. Other vehicle parameters are listed in TABLE ~\ref{TB:2}.

\begin{table}[!t]
 \renewcommand{\arraystretch}{1.3}
 \caption{Ego-vehicle parameters.}  \label{TB:2}
 \label{Table1}
 \centering
 \begin{tabular}{|c|c|c|}
 \hline
 {{\rm{Item}}} & Value & {{\rm{Description}}} \\
  \hline
 ${{m_T}}$ & {1300.0 {{\rm{kg}}}} & {{\rm{Total mass}}}  \\
 \hline
  ${{I_T}}$ & ${{{10}^4}} {{\rm{kg}} \cdot {{\rm{m}}^2}}$ & {{\rm{Moment of inertia}}}  \\
  \hline
  ${{K}}$ & 91090 & {{\rm{Tires stiffness}}}  \\
   \hline
  $\alpha_{T, 0}$ & -0.2 rad & {{\rm{\text { Initial vehicle sideslip angle }}}}  \\
 \hline
   $\mu $ & 0.8 & {{\rm{Friction coefficient}}}  \\
  \hline
  $\dot{\Psi}_{0} $ & 0.7 $rad/s$ & {{\rm{\text {Initial yaw rate }}}}  \\
    \hline
  ${v}_{0} $ & 16.75 $m/s$ & {{\rm{\text {Initial velocity }}}}  \\
 \hline
 \end{tabular}
 \end{table}


The ego-vehicle  state vector is ${S}_a = [X, Y, \Psi, \alpha, \dot {\Psi}]^{\top} \in \mathbb{R}^{5}$, its observable state vector is ${ S}_o = \{X,Y\} \in \mathbb{R}^{2}$, and its control input is $\delta$ steering angle. 

For the low-level controller, linear quadratic regulator (LQR) state feedback controller is utilized. To this end, given a planned trajectory ${\cal T}_p(t)$ based on the planned waypoints $(X_p,Y_p) \in {\cal T}_p(t)$ provided by high-level risk-averse $Q$-learning planner, the main objective of the low-level LQR state feedback controller is to follow ${\cal T}_p(t)$ and stabilize the ego-vehicle yaw, yaw rate, and side slip angle states, i.e., $s_v=[\Psi, \alpha, \dot {\Psi}]^{\top}$. That is,  $e(t) := ([ Y-Y_p,  \Psi, \alpha, \dot {\Psi}]^{\top})^2 \to 0$. To this aim, the LQR performance index is chosen as
$\int_{0}^{\infty}\left({{e}}^{T} \mathbf{Q}  {{e}}+{\delta }^{T} \mathbf{R}  {\delta }\right) \mathrm{d} t$ where $\mathbf{Q} = diag(3,1,1,1)$ and $\mathbf{R} = 1$. Note that $X-X_p$ is removed from $e(t)$ since $v_T=0$ and the low-level controller control the lateral dynamics of the ego-vehicle.

\begin{figure}[!t]
 \centering{\includegraphics [width=3.6in,height=2.2cm] {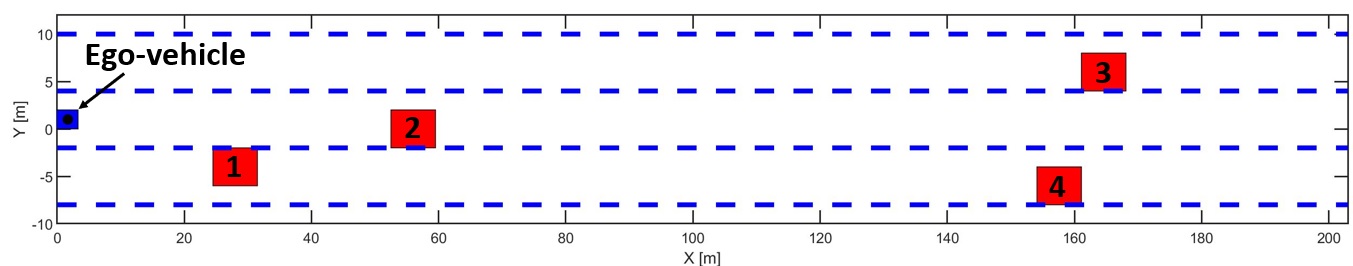}}
  \caption{Three-lane highway where the occupancy positions of the other vehicles are shown with red squares and ego-vehicle is shown with blue squares. Lanes are shown by dashed blue lines.}
  \label{FigSck}
\end{figure}

The ego-vehicle and the vehicles $2$, $3$, and $4$ speeds are $16.7 \mathrm{~m} / \mathrm{s}$, $2.6 m / s$, $-2.6 m / s$, and $-2.6 m / s$, respectively. In performing closed-loop experiments, the centers of gravity of ego vehicle and other vehicles are initialized at $(x(0), y(0)) = (0, 1) m$, $(x_1(0), y_1(0)) = (28, -4) m$, $(x_2(0), y_2(0)) = (56, 0) m$, $(x_3(0), y_3(0)) = (164.5, 6) m$, and $(x_4(0), y_4(0)) = (157.5, -6) m$. 
Fig.~\ref{FigSck} represents a $3$-lanes highway in which the positions of the ego-vehicle and other vehicles are shown with blue and red squares, respectively.

\begin{figure*}
  \includegraphics[width=\textwidth,height=8.4cm]{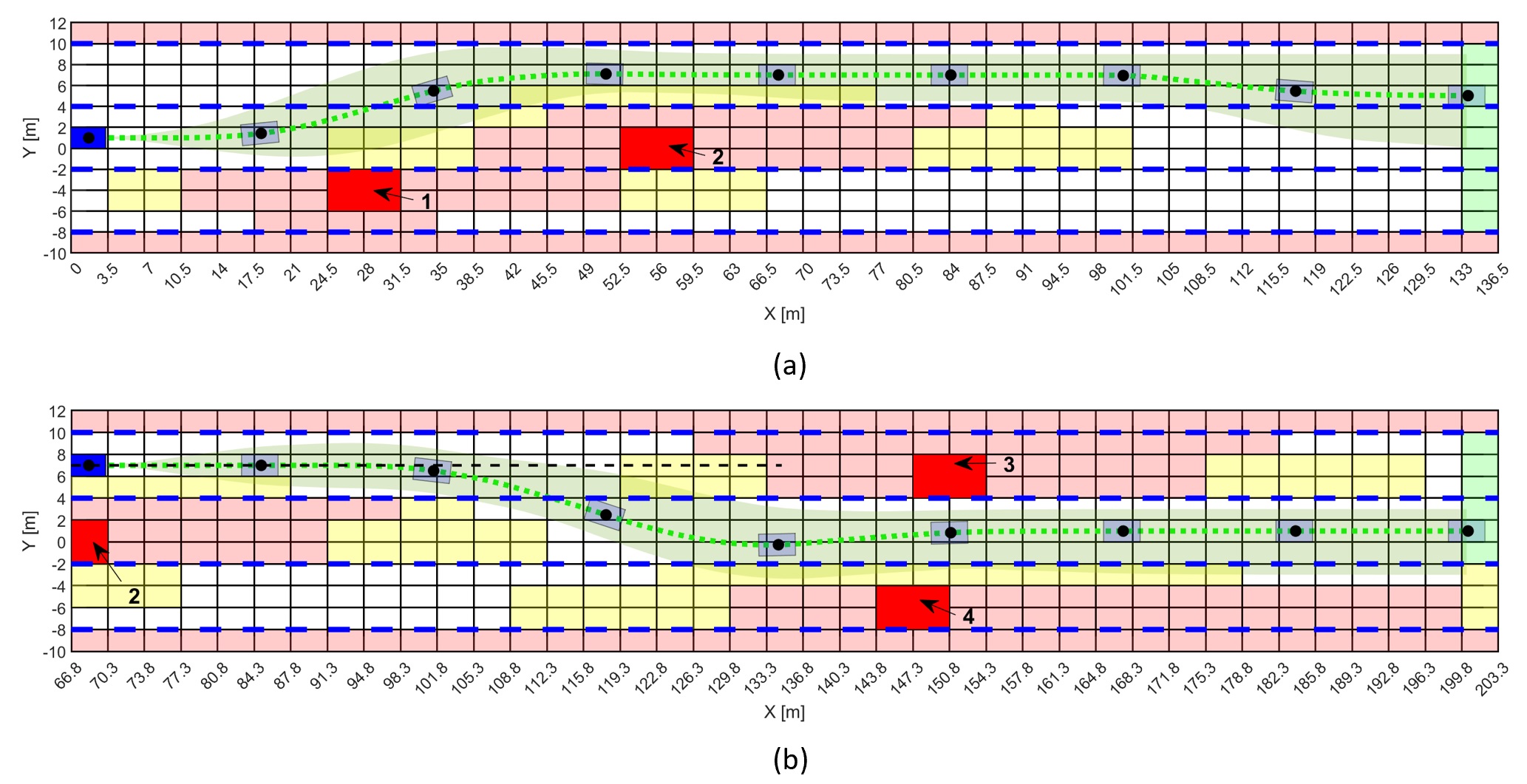}
  \caption{Three-lane highway represented by $11\times 39$ grid. Obstacles are red. Ego-vehicle is blue. Lanes are shown by dashed blue lines. The  shaded green area  and  dashed green line  represent  the  results between $[10\%,90\%]$ quantiles and the means across 10 independent experiments. }
  \label{FigSc1}
\end{figure*}

\begin{figure*}
  \includegraphics[width=\textwidth,height=3.3cm]{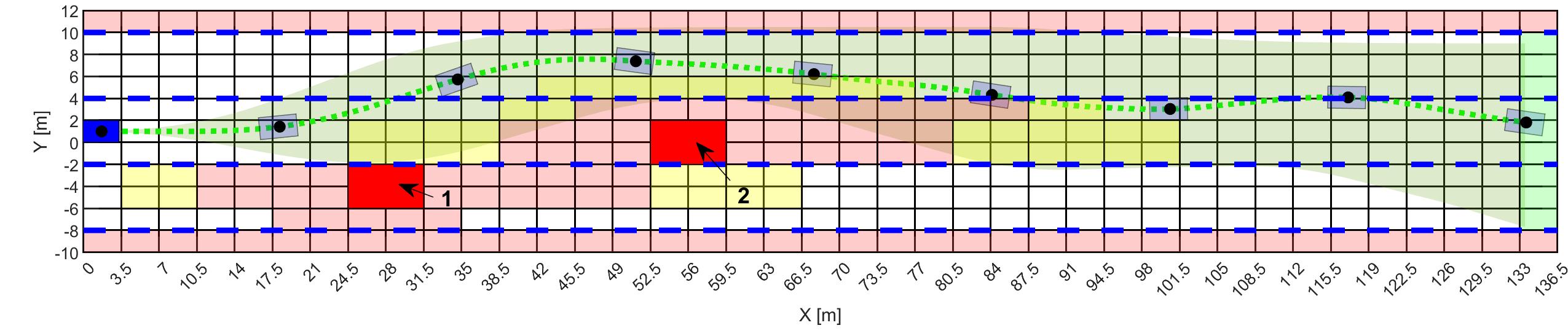}
  \caption{Risk Neutral case.}
  \label{FigSc1N}
\end{figure*}

The width and length of the highway are $22 m$ and $203 m$, respectively. We assume that the preview information is just available for the next eight seconds of travel. 
 { The environment is discretized and modeled as a  $2$-dimensional $11 \times 39$ grid with a total of $429$ cells with a width of $2 m$ where each square forward represents the next $0.2$ second of the travel (i.e., length of $3.5 m$ since the ego-vehicle traveling with the constant velocity of $16.75 m/s$ ) and cells can have different characteristics.} The dimensions of the ego-vehicle are considered as $2m \times 3.5 m$, and the dimensions of the other vehicles are chosen as $4 m \times 7 m$. Each of these cells in the environment can be in one of the five discrete states ${\cal Z}=\{un,sa,hr,lr,tg,cp\}$.
 {
Moreover, for simplicity, it is assumed that the evolution of stochastic occupancy of the road by other vehicles is available, i.e., the risk probabilities of the cells.
} 
The edges of the highway occupy one square, and each of the three lanes occupies three squares.  The finite set of high-level actions is ${\cal A} = \left\{ {\text{go straight,~left turn,~right turn}} \right\}$.


 
 
The simulation time is $12$ seconds, all the computations are carried out on an i7 1.9GHz Intel core processor with 8 GB of RAM running MATLAB R2021a, and the following values are used: $\gamma=0.3$, $\tau_{env}  \simeq  0.2 sec$, $n= 39$, and $N_k=1500$, $\tau = 15$, $M=-10000$, $\tau_{l}= 10000$, $\tau_{h}= 0$, $\sigma = 1$,  $\bar \tau_{l}= 10$, $\bar \tau{h}= 0$, $\bar \sigma = 1$, $\alpha = 0.2$.

\begin{figure}[!ht]
\centering{\includegraphics [width=3.4in] {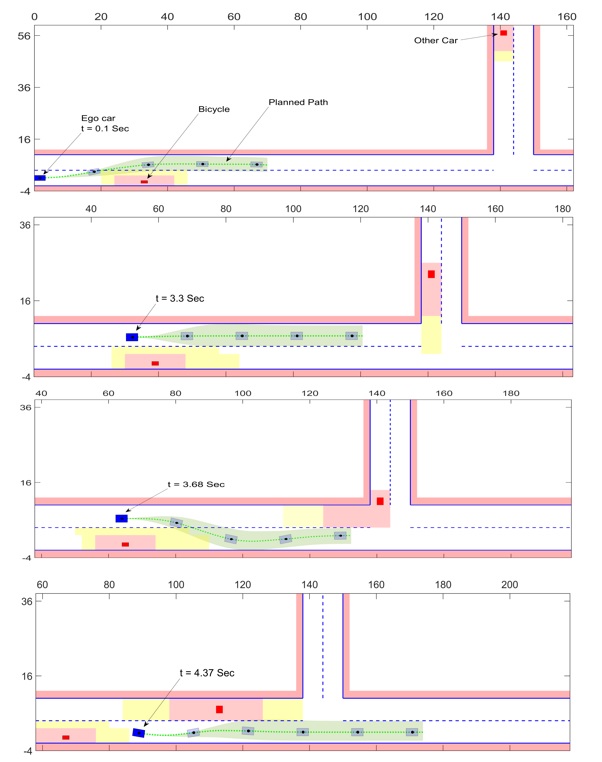}}
\caption{Illustration of the over-taking scenario. Ego-vehicle is blue. Lanes are shown by dashed blue lines. The  shaded green area  and  dashed green line  represent  the  results between $[10\%,90\%]$ quantiles and the means across ten independent experiments.} \label{Scenario2}
\end{figure}

The  trajectories  of  $X$ and $Y$ positions of the ego-vehicle for  both  risk-neutral (i.e., $\alpha = 0$)  and  risk-averse (i.e., $\alpha > 0$) cases  are  displayed  in  Figs.~\ref{FigSc1}. Fig.~\ref{FigSc1} represents two different time segments of the simulation, $[0 sec \sim  4 sec]$ and  $[4sec \sim 12sec]$. The  shaded green area  and  dashed green line  represent  the  results of the low-level continuous-time trajectories based on the high-level actions towards the terminal goal between $[10\%,90\%]$ quantiles and the means across 10 independent experiments. 
The successive frames of the ego vehicle are shown by the blue square to represent the current position of the ego vehicle and the transparent cyan squares to denote some of the planned estimated positions. Moreover, the current positions of the other vehicles are shown with red squares, and high-risk zones and low-risk zones are shown with transparent red and yellow squares, respectively, in the grids.  Note that at the time of $4~sec$, the FCU module detects the infeasibility issue and initiates new planning through the high-level risk-averse $Q$- learning.


Fig.~\ref{FigSc1N} represents the risk-neutral (i.e., $\alpha = 0$) planning for the first time segment of the simulation, i.e.,  $[0 sec \sim  4 sec]$. The  shaded green area  and  dashed green line  represent  the  results of the low-level continuous-time trajectories based on the high-level actions towards the terminal goal between $[10\%,90\%]$ quantiles and the means across 10 independent experiments. Note  that  the  variance  of  the planning position is significantly lower for the risk-averse case as expected.

 {
\textbf{Example 2}. In the second example, which is the overtaking scenario (which is  adopted from \cite{althoff2009model}), the main objective of ego-vehicle is to safely overtake a bicycle before reaching a T-intersection, where another car is approaching.
Using $s$ to represent the volumetric center of the vehicle along a path, $v$ to represent velocity, and $a$ to represent absolute acceleration, the longitudinal dynamics can be described as follows:
\begin{align}
    \dot{s}=v \quad \dot{v}= \begin{cases}5 \cdot\left(1-\left(v / 20\right)^{2}\right) \cdot u, & u>0 \\ 5 \cdot u, & u \leq 0 \\ 0, & v \leq 0\end{cases}
\end{align}
subject to the constraint $a \leq 5$ where $a=\sqrt{a_{N}^{2}+a_{T}^{2}},~ a_{N}=v^{2} / \rho(s)$,  $a_{T}=\dot{v}$, $\rho(s)$ maps the path coordinate $s$ to its radius of curvature, and $a_{N}$ and $a_{T}$, respectively, are the normal and tangential accelerations.
 We assume that the preview information is just available for the next three seconds of travel. The dimensions of the ego-vehicle are considered as $2m \times 3.5 m$, and the dimensions of the bicycle and other vehicle are chosen as $1 m \times 2 m$ and $2 m \times 2 m$, respectively.
The environment of the ego-vehicle is discretized according to \cite{althoff2009model}.
 The stochastic occupancy of the road by other vehicle are computed using probabilistic reachability analysis presented in \cite{althoff2009model}, which is online computationally efficient since most of the intensive computations are performed offline. 
Using this scenario, it is checked if an ego car can safely overtake a bicycle while another car approaches a T-intersection. 
Four time-segments of the simulation 
along with the risky zones (result from stochastic reachable sub-module) of the traffic participants are given in Fig.~\ref{Scenario2}.
The simulation time is $10$ seconds, and the following values are used: $\gamma=0.22$, $\tau_{env}  =  0.45 ~sec$, $v(0)=5 ~ m/s$, $N_k=1000$, $\tau = 10$, and $\alpha = 0.25$.
As mentioned, high-risk cells indicate high probability risk of collision, whereas low-risk cells represent areas of low probability risk of collision. 
Using the parallel computing toolbox (i.e., parfor), the total computational time is  $1.719$ Seconds, and the summation of computational times of the RAU and FCU modules are  $0.581$ and $0.306$ Seconds, respectively. With a similar setup, the total computational time of the method given in \cite{althoff2009model} is $1.481$ Seconds which indicates that it is less computationally demanding. However, the path provided by \cite{althoff2009model} is not optimal, and also, the proposed framework in this paper has the potential to provide us with the ability to transfer knowledge to other circumstances (e.g., by the similar idea provided by \cite{chen2022transferred,vallon2020data}), which can decrease the computational burden significantly.}

\section{Conclusion}

A risk-averse planner is developed for navigation of autonomous vehicles between lanes around moving vehicles. The multi-lane road ahead of a vehicle is represented by a finite-state MDP and the entire planning horizon is modeled as a series of MDPs that evolve over time. A risk assessment unit module is leveraged to assess and preview the risk of each state and the reward function. A static convex program-based preview-based $Q$ learning algorithm is then developed that leverages these probabilistic reward values to learn risk-averse optimal planning strategies. 
Finally, the developed scheme is implemented in some driving scenarios to show the efficiency of the proposed learning framework.
 {Using state-of-the-art vehicle simulators, such as CARLA, we intend to investigate this paper's approach with the ability of transferring knowledge in more depth in the future.}


%

\appendices

\section{Proof of Proposition \ref{prop:1}}

 To show the monotonicity property for $\hat{\mathbb{T}}$, one needs to show that for any ${Q_k^1}\left( {s,a} \right)$ and ${Q_k^2}\left( {s,a} \right)$, if ${Q_k^1}\left( {s,a} \right) \leq {Q_k^2}\left( {s,a} \right)$,  $\forall (s,a) \in S_k \times {\cal A}$ then $\hat{\mathbb{T}} {Q_k^1}\left( {s,a} \right) \leq \hat{\mathbb{T}} {Q_k^2}\left( {s,a} \right), \forall (s,a,a^\prime) \in S_k \times {\cal A}^2$. 
Now, consider ${Q_k^1}\left( {s,a} \right)$ and ${Q_k^2}\left( {s,a} \right)$. Let ${s_a^\prime }$  be the next state and $s$  be the current state. Since $\log (.)$ and $\exp (.)$ are injective relations (i.e., one-to-one and left-total) on  ${\mathbb{R}_{ > 0}}$, it follows from  ${Q_k^1}\left( {s_a^\prime ,a^\prime} \right) \le {Q_k^2}\left( {s_a^\prime ,a^\prime} \right)$,  that
\small
\begin{align}
\begin{array}{l}
\frac{1}{\alpha }\big( {\log {\mathbb{E}_{s_{a}^\prime \sim {\bar P}_{s,s^\prime}^{a}}}\big[ {\exp \big( {\alpha \gamma  {Q_k^1}\left( {s_a^\prime,a^\prime )} \right)} \big)} } \big] \le \\
\quad \quad 
\frac{1}{\alpha }\big( {\log {\mathbb{E}_{s_{a}^\prime  \sim {\bar P}_{s,s^\prime}^{a}}}\big[ {\exp \big( {\alpha \gamma  {Q_k^2}\left( {s_a^\prime,a^\prime} \right)} )} \big)} \big].
\end{array} \label{eq:28}
\end{align}
\normalsize
Adding ${g_a^k}\left( s \right)$  to both sides yields
\small
\begin{align}
\begin{array}{l}
{g_a^k}\left( s \right) +\frac{1}{\alpha }\big( {\log {\mathbb{E}_{s_{s}^\prime \sim {\bar P}_{s,s^\prime}^{a}}}\big[ {\exp \big( {\alpha \gamma  {Q_k^1}\left( {s_a^\prime,a^\prime )} \right)} \big)} } \big] \le 
\\ \qquad
{g_a^k}\left( s \right) +\frac{1}{\alpha }\big( {\log {\mathbb{E}_{s_{a}^\prime  \sim {\bar P}_{s,s^\prime}^{a}}}\big[ {\exp \big( {\alpha \gamma  {Q_k^2}\left( {s_a^\prime,a^\prime )} \right)} \big)} } \big],
\end{array}
\label{eq:29}
\end{align}
\normalsize
which implies that  $\widehat {\mathbb{T}}{Q_k^1}\left( {s,a} \right) \le \widehat { \mathbb{T}}{Q_k^2}\left( {s,a} \right),\forall (s,a,a^\prime) \in {S_k} \times {\cal A}^2$. Therefore,
\begin{align}
\left\langle {\widehat {\mathbb{T}}{Q_k^1}\left( {s,a} \right) - \widehat {\mathbb{T}}{Q_k^2}\left( {s,a} \right),{Q_k^1}\left( {s,a} \right) - {Q_k^2}\left( {s,a} \right)} \right\rangle  \ge 0, \label{eq:30}
\end{align}
 {$\forall {Q_k^1},~{Q_k^2} \in {\cal C} ({S_k} \times {\cal A}) $, which satisfies the condition of Definition \ref{defn:4}.} This completes the proof.\hfill $\square$

\section{Proof of Theorem \ref{theorem:1}}
Before proceeding with the proof, note that using Definition \ref{defn:5}, one has 
\small
\begin{align}  \label{eq:31}
\begin{array}{l}
\qquad \alpha \gamma ({Q_k^2}( {s , a} ) -  
{Q_k^1}( {s,a}) ) 
\\
\qquad \qquad \le \alpha \gamma \big\| {Q_k^1}( {s,a} ) 
 - {Q_k^2}\left( {s,a} \right) \big\|_w w(s,a) \\
 \Leftrightarrow 
 \alpha \gamma {Q^{{{\pi_k^2}}}}\left( {s,a} \right) - 
 \\
 \qquad  \alpha \gamma \big\|  {Q_k^1}\left( {s,a} \right)  -  {Q^{{{\pi_k^2}}}}\left( {s,a} \right){\big\|_w}w(s,a) 
 \le \alpha \gamma {Q_k^1}\left( {s,a} \right) 
 \\
   \Leftrightarrow 
   \log {\mathbb{E}_{s_{a}^\prime \sim {\bar P}_{s,s^\prime}^{a}}}\big[ \exp \big( 
    \alpha \gamma {Q_k^2}\left( {s,a} \right) -
    \\
 \qquad  \alpha \gamma \big\|  {Q_k^1}\left( {s,a} \right)  -  {Q_k^2}\left( {s,a} \right){\big\|_w}w(s,a)
  \big) \big] 
  \\
\qquad \qquad \le \log {\mathbb{E}_{s_{a}^\prime \sim  {\bar P}_{s,s^\prime}^{a}}}\big[ \exp \big( 
 \alpha \gamma {Q_k^1}\left( {s,a} \right)
 \big) \big]  
 \\
  \Leftrightarrow
   \frac{{ - 1}}{\alpha } \log {\mathbb{E}_{s_{a}^\prime \sim {\bar P}_{s,s^\prime}^{a}}}\big[ \exp \big( 
    \alpha \gamma {Q_k^2}\left( {s,a} \right) - 
    \\
  \qquad \alpha \gamma \big\|  {Q_k^1}\left( {s,a} \right)  -  {Q_k^2}\left( {s,a} \right){\big\|_w}w(s,a)\big) \big] 
  \\
\qquad \qquad \ge \frac{{ - 1}}{\alpha }
 \log {\mathbb{E}_{s_{a}^\prime \sim  {\bar P}_{s,s^\prime}^{a}}}\big[ \exp \big( 
 \alpha \gamma {Q_k^1}\left( {s,a} \right)) \big] 
  \end{array}
\end{align}
\normalsize
Then,
\small
\begin{align}
\begin{array}{l}
 \widehat {\mathbb{T}}{Q_k^2}(s,a) - \widehat {\mathbb{T}}{Q_k^1}(s,a) \le
\\
\mathop {\sup }\limits_{a^\prime \in {\cal A}} \big( - \frac{1}{\alpha }\log {\mathbb{E}_{ s_{a}^\prime \sim {\bar P_{s,{s^\prime }}^{a}}}}[\exp (\alpha \gamma {Q_k^1}(s_{a}^\prime ,a^\prime))]
\\
\qquad  + \frac{1}{\alpha } \log {\mathbb{E}_{s_{a}^\prime \sim  {\bar P_{s,{s^\prime }}^{a}} }}[\exp (\alpha \gamma  {Q_k^2}(s_{a}^\prime ,a^\prime))]\big) 
 \le  \\
 \mathop {\sup }\limits_{a^\prime \in {\cal A}} \big(  - \frac{1}{\alpha }\log {\mathbb{E}_{s_{a}^\prime \sim  {\bar P}_{s,s^\prime}^{a} }}\big[ \exp \big( \alpha \gamma \times \\
\quad 
\big\| {Q_k^1}\big( {s_{{a}}^\prime ,a^\prime} \big)   - {Q_k^2} \big( {s_{{a}}^\prime ,a^\prime} \big) \big\|_w w\big( {s_a^\prime }, a^\prime \big) +
 \alpha \gamma {Q_k^2}\big( {s_{a}^\prime ,a^\prime )} \big) \big) \big]  
 \\
  + \frac{1}{\alpha }\log {\mathbb{E}_{s_a^\prime \sim {\bar P}_{s,s^\prime}^{a}}}\big[ {\exp \big( \alpha \gamma 
  {Q_k^2}\big( {s_a^\prime ,a^\prime )} \big)} \big] \big) \le
  \\
  \mathop {\sup }\limits_{a^\prime \in {\cal A}} \big(  - \frac{1}{\alpha }\log {\mathbb{E}_{s_a^\prime \sim {\bar P}_{s,s^\prime}^{a}}}\big[ \exp \big( \alpha \gamma \big\| 
 {Q_k^1}\big( {s_a^\prime ,a^\prime )} \big) -\\
   {Q_k^2}\big( {s_a^\prime ,a^\prime )} \big) \big\|_w w\big( {s_a^\prime, a^\prime } \big) \big] -
   \frac{1}{\alpha }\log {\mathbb{E}_{s_a^\prime \sim {\bar P}_{s,s^\prime}^a}}\big[ {\exp \big( \alpha \gamma 
  {Q_k^2}\big( {s_a^\prime ,a^\prime )} \big)} \big]
  \\
 + \frac{1}{\alpha }\log {\mathbb{E}_{s_a^\prime \sim  {\bar P}_{s,s^\prime}^a }}\big[ {\exp \big( \alpha \gamma 
{Q_k^2}\big( {s_a^\prime ,a^\prime )} \big)} \big] \big)
 \\
 = \mathop {\sup }\limits_{a^\prime \in {\cal A}} \big( \frac{-1}{\alpha }\log {\mathbb{E}_{s_a^\prime \sim {\bar P}_{s,s^\prime}^a}}\big[ \exp \big( \alpha \gamma 
 \big\| {Q_k^1}\big( {s_a^\prime ,a^\prime )} \big)
 \\ \qquad 
 -  {Q_k^2}\big( {s_a^\prime ,a^\prime )} \big) \big\|_w w\big( {s_a^\prime, a^\prime } \big) \big] \big) 
 \\
  \le \mathop {\sup }\limits_{a^\prime \in {\cal A}} \big(  - \frac{1}{\alpha }{\mathbb{E}_{s_a^\prime \sim {\bar P}_{s,s^\prime}^a}}\log \big[ \exp \big( \alpha \gamma \times \\
 \qquad  \big\| {Q_k^1}\big( {s_a^\prime ,a^\prime )} \big) 
 -  {Q_k^2}\big( {s_a^\prime ,a^\prime )} \big) \big\|_w w\big( {s_a^\prime, a^\prime } \big) \big] \big)
 \\
 = \mathop {\sup }\limits_{a^\prime \in {\cal A}} \big( \frac{1}{\alpha }{\mathbb{E}_{s_a^\prime \sim  {\bar P}_{s,s^\prime}^a}}\big( \alpha \gamma \big\| 
 {Q_k^1}\big( {s_a^\prime ,a^\prime )} \big)
 -  {Q_k^2}\big( {s_a^\prime ,a^\prime )} \big) \big\|_w w\big( {s_a^\prime, a^\prime } \big) \big) \big),
\end{array} \label{eq:33Proof}
\end{align}
\normalsize
where the second inequality follows from \eqref{eq:31}, the third inequality from  $\log \mathbb{E}[\bar f(s) + \bar g(s)] \ge \mathbb{E}(\bar f(s)) + \mathbb{E}(\bar g(s))$, and the last inequality from Jensen’s inequality. It follows from \eqref{eq:33Proof} and Assumption \ref{Assm:2} that
\small
\begin{align}
\begin{array}{l}
    \big\| \widehat {\mathbb{T}}{Q_k^2}(s,a) - \widehat {\mathbb{T}}{Q_k^1}(s,a)\big\|_w w\big( {s, a } \big) \le \\
     \mathop {\sup }\limits_{s_a^\prime \in {S_k}} | \mathop {\sup }\limits_{a^\prime \in {\cal A}} \big( \gamma {\mathbb{E}_{s_a^\prime \sim  {\bar P}_{s,s^\prime}^a}}\big( \big\|  {Q_k^1}  -  {Q_k^2} \big\|_w w\big( {s_a^\prime, a^\prime } \big) \big) \big)|\\
 \quad \quad \quad\quad \quad \quad\quad \quad \quad\quad \quad \quad\quad \quad \le \Upsilon \gamma {\big\| {{Q_k^1} - {Q_k^2}} \big\|_w}w\big( {s, a } \big),
 \end{array} 
\end{align}
\normalsize
 which implies \eqref{eq:22}. This completes the proof.  \hfill $\square$

\section{Proof of Corollary \ref{Corollary:1} }
Recall that
\small
\begin{align}
\begin{array}{l}
{g_a^k}\left( s \right) + \frac{1}{\alpha } \log \big({\mathbb{E}_{s_a^\prime \sim {\bar P}_{s,s^\prime}^a}}\big[ \exp \big( \alpha \gamma 
  \mathop {\inf }\limits_{a^\prime \in {{\cal A}}} \big\{{Q_k^{*}}\left( {s_a^\prime ,a^\prime} \right) \big\} \big) \big] \big)   
  \le 
  \\ {g_u}\left( s \right) + \frac{1}{\alpha } 
  \log \big({\mathbb{E}_{s_a^\prime \sim {\bar P}_{s,s^\prime}^a}}\big[ {\exp \big( {\alpha \gamma 
  {Q_k^{*}}\left( {s_a^\prime ,a^\prime} \right)} )} \big] \big)
\end{array} \label{eq:33}
\end{align}
\normalsize
$\forall (s,a,{a}^\prime ) \in {S_k} \times {\cal A}^2  $. This implies that  ${Q_k^{*}}$ is a feasible solution to \eqref{eq:25}. Also, for any $Q_k \in {\cal C} ({ S_k} \times {\cal A})$  in a feasible region of \eqref{eq:25}, the inequality 
\small
\begin{align}
\begin{array}{l}
Q_k(s,a)\, \le {g_a^k}\left( s \right) + \frac{1}{\alpha } 
\log {\mathbb{E}_{s_a^\prime \sim {\bar P}_{s,s^\prime}^a}}\big[ {\exp \big( {\alpha \gamma Q( {s_a^\prime,a^\prime} )} \big)} \big]
\end{array} \label{eq:34}
\end{align}
\normalsize
holds 
$\forall (s,a,{a}^\prime) \in {S_k} \times {\cal A}^2 $, which implies that it also holds for specific $a^\prime \in {\cal A}$  that minimizes  $Q_k\left( {s_a^\prime ,{a^\prime }} \right)$. That is, it also satisfies $ {\mathbb{T}}$  operator, i.e.,  $Q_k(s,a) \le  {\mathbb{T}}Q_k(s,a) = {g_a^k}\left( s \right) + \frac{1}{\alpha } \log \big({\mathbb{E}_{s_a^\prime \sim {\bar P}_{s,s^\prime}^a}}\big[ \exp \big( \alpha \gamma 
  \mathop {\inf }_{a^\prime \in {{\cal A}}} \big\{{Q_k^{*}}\left( {s_a^\prime ,a^\prime} \right) \big\} \big) \big] \big)$. Based on Proposition \ref{prop:3}, this indicates that $Q_k$ is a lower bound to  ${Q_k^{*}}$. Since we are maximizing the objective function in \eqref{eq:25}, this means that ${Q_k^{*}}$  is a maximizer of the optimization. This completes the proof. 
\hfill $\square$



\ifCLASSOPTIONcaptionsoff
  \newpage
\fi



%

\bibliographystyle{IEEEtran}
\bibliography{ref}




%






\vspace{-1cm}
\begin{IEEEbiography}[{\includegraphics[width=1in,height=1.25in,clip,keepaspectratio]{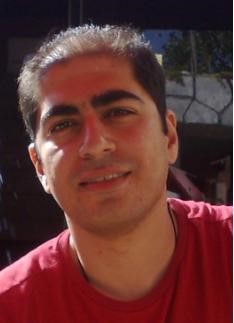}}]{Majid Mazouchi} 
 received his B.Sc. degree in Electrical Engineering from K. N. Toosi University of Technology, Iran, in 2007 and his M.Sc. and Ph.D. degrees in Electrical Engineering from Ferdowsi University of Mashhad, Iran, in 2010 and 2018, respectively. He was a Senior Lecturer at Semnan University, from 2017 to 2018. He is currently a postdoctoral research fellow in Mechanical Engineering Department, Michigan State University, East Lansing, MI, USA. His current research interests include multi-agent systems, reinforcement learning, cyber-physical systems, and distributed control.
\end{IEEEbiography}

\vspace{-1cm}
\begin{IEEEbiography}[{\includegraphics[width=1in,height=1.25in,clip,keepaspectratio]{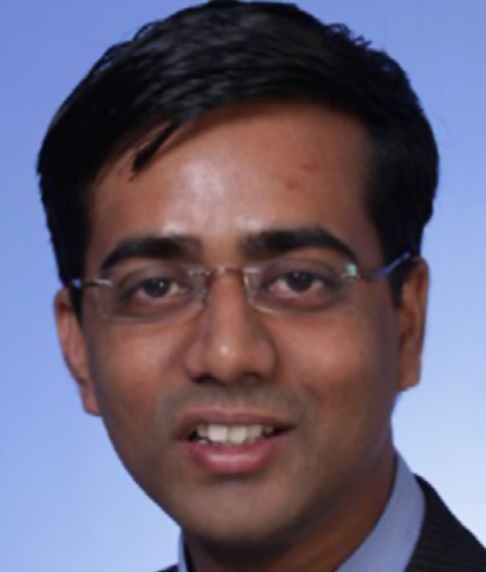}}]{Subramanya P. Nageshrao} 
received the B.E. degree in electronics and communication from Visvesvaraya
Technological University, Belgaum, India, in 2005, M.S. degree in mechatronics from Technische Universität Hamburg-Harburg, Hamburg, Germany, in 2011, and the
Ph.D. degree from Delft University of Technology, Delft, The Netherlands in 2016.
From 2005 to 2009, he was a Software Engineer with Bosch Ltd., Bangalore, India. His current research interests include nonlinear control, distributed control,
machine learning, particularly reinforcement learning and its applications for mechatronic and robotic systems.
\end{IEEEbiography}
\vspace{-1cm}

\begin{IEEEbiography}[{\includegraphics[width=1in,height=1.25in,clip,keepaspectratio]{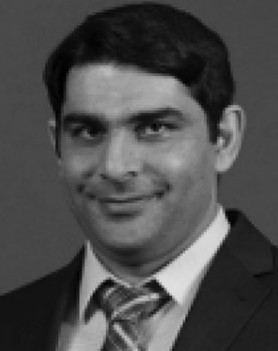}}]{Hamidreza Modares}  (M'15) received the B.S. degree from the University of Tehran, Tehran, Iran, in 2004, the M.S. degree from the Shahrood University of Technology, Shahrood, Iran, in 2006, and the Ph.D. degree from the University of Texas at Arlington, Arlington, TX, USA, in 2015. He was a Senior Lecturer with the Shahrood University of Technology, from 2006 to 2009 and a Faculty Research Associate with the University of Texas at Arlington, Arlington, TX, USA from 2015 to 2016. He is currently an Assistant Professor in Mechanical Engineering Department, Michigan State University, East Lansing, MI, USA. His current research interests include cyber-physical systems, reinforcement learning, distributed control, robotics, and machine learning. Dr. Modares is an Associate Editor for the IEEE TRANSACTIONS ON NEURAL NETWORKS AND LEARNING SYSTEMS. He was the recipient of the best paper award from 2015 IEEE International Symposium on Resilient Control Systems.
\end{IEEEbiography}

\vfill



\end{document}